\documentclass[review]{elsarticle_nonatbib}

\usepackage{lineno,hyperref}
\modulolinenumbers[5]

\journal{ArXiv}

\usepackage{xargs}                      
\usepackage[pdftex,dvipsnames]{xcolor}  
\usepackage[disable, colorinlistoftodos,prependcaption,textsize=footnotesize]{todonotes}

\usepackage{amsmath}
\usepackage{amssymb}
\usepackage{hyperref}       
\hypersetup{
	colorlinks=true,
	linkcolor=blue,
	filecolor=magenta,      
	urlcolor=cyan,
	citecolor=blue
}

\usepackage{amsfonts}       
\usepackage{microtype}      
\usepackage{graphicx}

\usepackage{enumitem}

\usepackage{algorithm} 
\usepackage{caption}
\DeclareCaptionFormat{algorithm}{{\parbox[c][1.75em][c]{\textwidth}{\hspace{0.25em}#1#2#3}}}
\newcommand*\dif{\mathop{}\!\mathrm{d}} 

\makeatletter
\newcommand{\Mc}{}
\DeclareRobustCommand{\Mc}{%
	M%
	\raisebox{\dimexpr\fontcharht\font`M-\height}{%
		\check@mathfonts\fontsize{\sf@size}{0}\selectfont
		{c}%
	}%
}
\makeatother

\usepackage{accents}
\newcommand{\myvect}[1]{\accentset{\rightharpoonup}{#1}}

\usepackage[backend=biber, 
			style=numeric, 
			url=false, 
			maxcitenames=2, 
			natbib=true, 
			sorting=none]{biblatex}
\usepackage{doi}

\begin{document}
\begin{frontmatter}

\title{An augmented Lagrangian algorithm for recovery of ice thickness in unidirectional flow using the Shallow Ice Approximation.}


\auth[math]{Elizabeth K. \Mc George}

\auth[math]{Miguel Moyers-Gonzalez\corref{corresponding-author}}
\ead{miguel.moyersgonzalez@canterbury.ac.nz}

\auth[math,tpm]{Phillip L. Wilson}

\auth[mech]{Mathieu Sellier}

\cortext[corresponding-author]{Corresponding author.}

\address[math]{Te Kura Pāngarau | School of Mathematics and Statistics, Level 4, Jack Erskine Building, University of Canterbury, Private Bag 4800, Christchurch 8140, New Zealand}

\address[tpm]{Te Pūnaha Matatini Centre of Research Excellence, University of Auckland, Auckland, New Zealand}

\address[mech]{Department of Mechanical Engineering, University of Canterbury, Christchurch, New Zealand}

\begin{abstract}

A key parameter in ice flow modelling is basal slipping at the ice-bed interface as it can have a large effect on the resultant ice thickness. Unfortunately, its contribution to surface observations can be hard to distinguish from that of bed undulations. 
Therefore, inferring the ice thickness from surface measurements is an interesting and non-trivial inverse problem.
This paper presents a method for recovering dually the ice thickness and the basal slip using only surface elevation and speed measurements.
The unidirectional shallow ice approximation is first implemented to model steady state ice flow for given bedrock and basal slip profiles. This surface is then taken as  synthetic observed data. An augmented Lagrangian algorithm is then used to find the diffusion coefficient which gives the best fit to observations. Combining this recovered diffusion with observed surface velocity, a simple Newton's method is used to recover both the ice thickness and basal slip.
The method was successful in each test case and 
this implies that it should be possible to recover both of these parameters in two-dimensional cases also.
\end{abstract}


\end{frontmatter}

\linenumbers

\section{Introduction}

Ice thickness recovery from surface data is a popular problem among those working on land ice models, with many different sectors all seeking to understand the intricacies of the problem. Governments need the information for policy and natural resource planning. Geo-scientists need more detailed resolution in the bed topography to fully understand glacial processes \cite{Morlighem2020}. Statisticians question how certain bed inversions can be given the uncertain nature of many factors in any model \cite{Raymond2009, Babaniyi2021}, and mathematicians wonder if the solution is even unique.

As stated in the Summary for Policymakers chapter of the 2019 IPCC Special Report on the Ocean and Cryosphere in a Changing Climate \cite{IPCC2019_policy}, all people on earth depend directly or indirectly on the ocean and cryosphere. The cryosphere refers to the frozen component of the earth system and includes land ice in the form of snow, glaciers, ice caps, permafrost, and ice sheets, as well as frozen parts of the ocean such as those surrounding Antarctica and Greenland. It also includes frozen lakes and rivers \cite{NOAA2019}. Oceans cover 71\% of the Earth surface and land ice covers approximately 10\% of Earth's land area.
Populations living in coastal environments and mountainous regions are particularly vulnerable to changes in ocean and cryosphere.
Around 680 million people live in in low lying coastal zones and another 670 in high mountain regions (totally approximate 20\% of the 2010 global population).  For these people, ocean and cryosphere provide life-sustaining services such as food and water supply, renewable energy, and benefits for health and well-being, cultural values, tourism, trade, and transport.
Given the potentially large impact of climate projections on human livelihood, comprehensive and accurate predictive models for ocean and cryosphere dynamics are needed to support policy planning in governments.

A key component of cryosphere dynamics is that of land ice.
The contribution of land ice to global mean sea level (GLMS) rise for medium emissions scenarios is projected to be at least 0.10 m by 2100 with some models predicting a contribution of up to 0.27 m \cite{Church2013}.
\citet{Church2013} identified one of the main contributors to this rise as the melting of land ice.
To track the evolution of ice mass, ice thickness measurements or estimations are needed. However, these are costly measurements to take over large areas.
Due to this, scientists often estimate ice thickness based on a few measurements or from surface data.
To make these estimations first requires some model of how the ice flows.

Ice sheets have two main characteristics; (1) they exhibit gravity-driven creep flow which is sustained by the underlying sloped geography and (2) their growth and/or decline is controlled by the accumulation and/or ablation due to snow fall and/or melting.
Ice can be categorised as an incompressible, nonlinear, viscous, heat conducting fluid \cite{Greve2009} and can be described mathematically by the full Navier-Stokes (NS) flow equations together with a Generalized Newtonian Constitutive Law (Glen's Law). Many methods of approximating the conservation equations for ice sheets have been proposed in the last century. These approximations are not all equal; each have their own advantages and drawbacks. Typically, the more simplistic a model, the faster and easier it is to use in computations\todo{CHECK: is this is what Miguel wanted changed?}. But, of course, these simple models can omit processes which are important for accurately capturing the flows' behaviour.

One of the most widely used approximations for ice sheet flow is the shallow-ice approximation (SIA) \cite{Hutter1981, Fowler1987}. 
The SIA simplifies the full Stokes equations by performing a scaling analysis to obtain simplified field equations for the ice sheet flow. This scaling assumes the ice extent is much larger than its thickness. \citet{Blatter2011} advise caution when applying the SIA to processes on smaller scales where the assumptions may no longer be valid, for example, anisotropic basal sliding or locally steep basal topography. 
Simply put, in the SIA model, gravity-driven ice flow is solely balanced by basal drag neglecting longitudinal and transverse stresses, as well as vertical stress gradients \cite{Adhikari2012}.
Despite potential drawbacks, the SIA is used widely in ice flow modelling as it reduces a three dimensional flow with four unknown fields into problem into a two dimensional problem sith a single unknown field. This makes it computationally simple in comparison to the full Stokes where a full force balance has to be calculated at each step.

Due to the complexity of ice flow behaviour, recovering the ice thickness from only surface measurements is a non-trivial inverse problem.
In current state of the art models, it has been shown that variations in recovered ice thickness can be as large as the ice thickness recovered. The recovered thickness is also very sensitive to input data \cite{Farinotti2017}.
These variations are due to, in part, placing excess assumption on the flow behaviour, such as the no-slip condition at the base \cite[][]{Barcilon1993, Wilchinsky2001, Adhikari2011, Gessese2015, Heining2016}.

Imposition of a no-slip condition simplifies the inverse problem, allowing much faster computation.
However, basal slip is known influence the flow behaviour \cite{Jiskoot2011} and should be included if possible.
Flow speed is modulated by the presence, or lack thereof, of friction at the ice-bedrock boundary \cite{Cuffey2010} as well as the steepness of the underlying slope.
Since the free surface of an ice flow is affected by both basal slip and bedrock topography, separating the effects of these two factors in the recovery is difficult \cite{Martin2015, Monnier2017a}.

\citet{Bueler2009} use the shallow-shelf approximation (SSA) as a sliding law for the shallow-ice approximation. The SSA, derived originally by \citet{Morland1987} and \citet{Macayeal1989}, assumes that basal shear stresses are negligible since the shelf is floating and so longitudinal stresses dominate. They hoped to bridge the gap between observations of varying velocity across ice sheets and modelled velocities. In their paper, they use an average of velocities from the shallow shelf approximation and the non-sliding shallow-ice approximation in the energy conservation and mass continuity equations. The resultant velocity field exhibits realistic behaviour as seen in observations of ice streams.
One way to try to decouple the effects of bed topography and basal slip is to assume some a priori knowledge of the particular ice flow. \citet{Zorzut2020} included basal slip in their ice thickness estimations for the Monte Tronador glaciers using the parallel flow approximation. The parallel flow approximation assumes that glaciers deform only by simple shear such that flow lines are parallel \cite{Cuffey2010}. To do this, they assumed a linear proportionality between basal speed and surface speed and used measured points of ice thickness to compute an estimation of the factor. 
Recent work towards understanding the coupled behaviour of bed topography and basal slip in the ice surface presentation is promising. \citet{Monnier2017a} take surface data together with an initial ice thickness estimate from measurements and then improve upon it. This is done by optimising an objective functional to match the SIA modelled surface and observations (sometimes called variational data assimilation or VDA). Adding to this, \citet{Monnier2019} consider an alternative form of the SIA, which combines all unknown parameters in one variable. Incorporated in this variable is allowance for basal slipping, however it is not explicitly recovered.

Further work is also being done towards understanding the sensitivity of inversions to their inputs. \citet{Kyrke-Smith2017} investigated whether there is any correlation between seismic observations of basal acoustic impedance and the basal slip obtained from surface-to-bed inversions. Using data from the Pine Island Glacier, they found that, on the kilometre scale, there was no significant correlation between acoustic impedance and basal slip. However, using averaged values along profiles on the glacier resulted in a stronger correlation causing them to hypothesise that there may be correlation over length scales which are important to overall ice sheet flow. This sort of research using available and easy to measure data  helps to restrict uncertainty in the basal slip parameter. \citet{Kyrke-Smith2018} further went on to consider the effect of of bed resolution in basal slip inversions using the Pine Island data.

Bedrock data is costly and time consuming to acquire and is not available in many cases. It is therefore useful to analyse the effect of the bed on basal slip inversions.
To do this, \citeauthor{Kyrke-Smith2018} look at the sensitivity of inversion methods to the accuracy of the bedrock profile prescribed. In particular, they consider the separated effect of basal drag due to the bed topography (form drag) and the drag due to bed properties (skin drag).
They showed that a significant amount of basal shear calculated in inversion may be due to unresolved bed topography.
This reinforces that an inversion model solving for both basal slip and bed topography could be used to overcome the shortcoming.

\citet{Cheng2020} explore the sensitivity of modelled free surfaces in ice flows to their basal topography and basal friction. They compare results from both the full Stokes model and the SSA. To compute the sensitivites, they use the adjoint equations to compute gradients for the perturbed data with respect to the basal conditions. They found that the sensitivity depended on the wavelength of the perturbation and its distance to the grounding line (the location where the ice sheet transitions from land to floating). As expected, changes in the topography can be directly seen in the surface whereas changes in the friction coefficient are more subtle.

An overview of the governing ice flow model used is given in Sect. \ref{sec:equations}. 
Section \ref{sec:inverse_methodology} goes through the construction of synthetic glacier surfaces for a number of different cases and then gives the methodology and algorithms needed for the inverse problem.
The results of implementing the inverse method are given in Sect. \ref{sec:numerical_results} and additionally, a brief sensitivity analysis of these results to noise in surface data is covered in Subsec. \ref{subsec:results_noise}.
Finally, the results are discussed in Sect. \ref{sec:discussion} and final conclusions drawn in Sect. \ref{sec:conclusion}.

\section{Governing equations}
\label{sec:equations}
Beginning from the full Stokes flow equations for an ice sheet, conservation of mass for an incompressible fluid gives
\begin{align}
\nabla \cdot \myvect{u} = 0
\end{align}
and conservation of momentum gives
\begin{align}
\rho \frac{D \myvect{u}}{ D t} &= -\nabla p + \nabla \cdot \boldsymbol{\tau} + \myvect{f}\
\end{align}
where $\frac{D}{Dt} \equiv \frac{\partial }{\partial t} + \myvect{u} \cdot \nabla$ is the material derivative,  $\myvect{u}$ is the flow velocity of the ice sheet, $\rho$ the density, $p$ the pressure, $\nabla \cdot$ the divergence, $\boldsymbol{\tau}$ the deviatoric stress tensor, and $\myvect{f}$ the body forces experienced by the ice sheet, namely gravity. 

Pairing these conservation equations with the tensorial form of Glen and Nye's rheological law to describe the relationship between strain and shear \cite{Glen1952, Nye1953}
\begin{align}
\label{eq:glen-tensor-form}
	\dot{\boldsymbol{\gamma}} &= A(T) |\boldsymbol{\tau}|^{\frac{n-1}{2}} \boldsymbol{\tau},
\end{align}
the conservation equations for ice flow in this paper (presented in Subsec. \ref{subsec:sia}) can be derived. Here $\dot{\boldsymbol{\gamma}}$ is the strain rate tensor, $A(T)$ is a temperature dependent constant, and $|{\boldsymbol{\tau}}|$ denotes the second invariant. Classically, the flow of ice is assumed to be well described for $n=3$.

\subsection{Shallow ice approximation (SIA)}
\label{subsec:sia}
\begin{figure}
	\centering
	\includegraphics[width=0.5\linewidth]{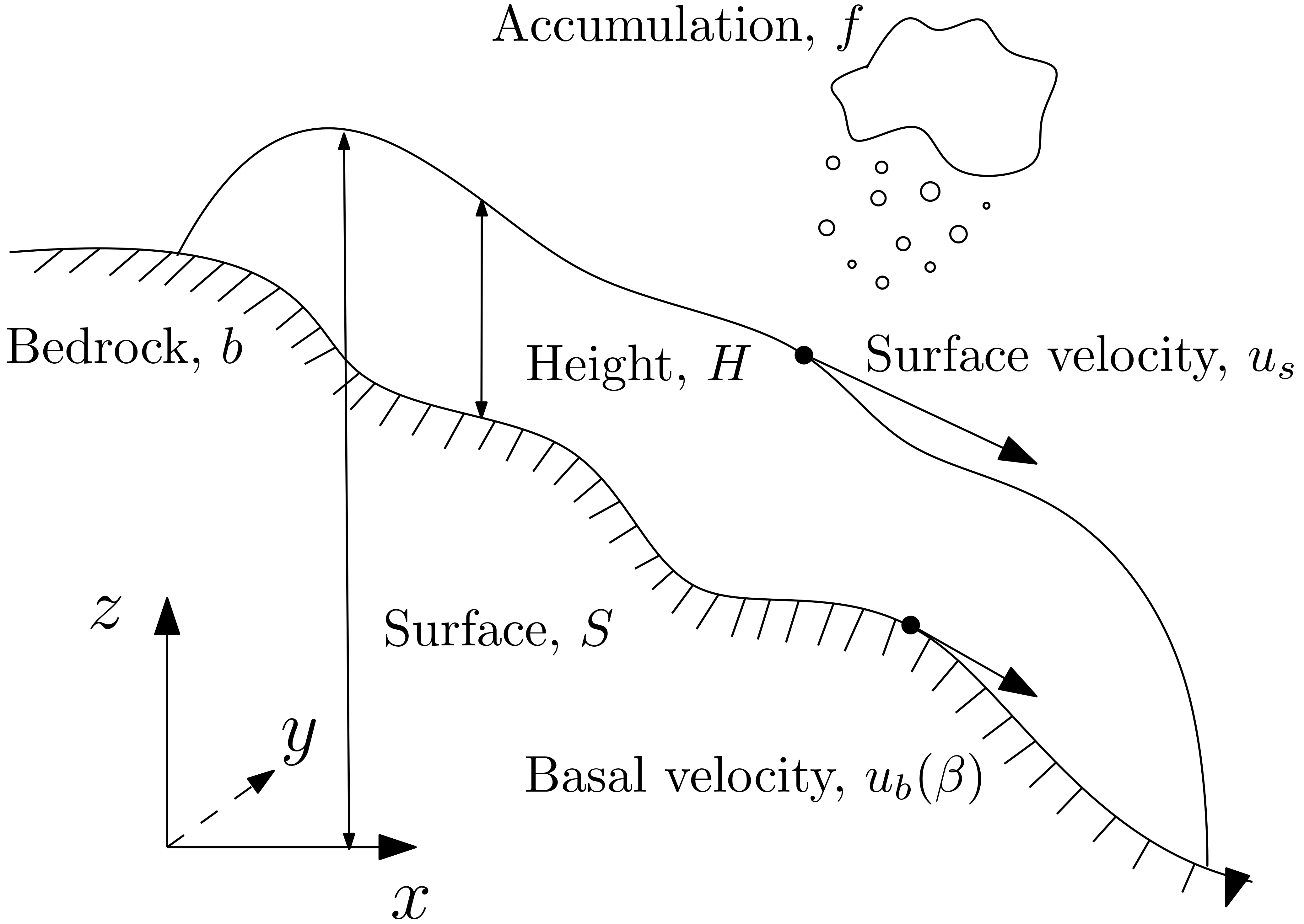}
	\caption{Ice sheet flowing downstream with surface $S$, bedrock $b$, and height $H$. The accumulation $f$ is indicated by falling precipitation. Surface speed, $u_s$, and basal slip, $\beta$, are also indicated.}
	\label{fig:variables}
\end{figure}

Under shallow-ice assumptions, the constitutive equations reduce to the SIA giving ice thickness evolution over time. From this thickness profile, the surface speed can be subsequently recovered. The SIA is chosen due to its relative simplicity. Typically the coordinate system is set up with the $x$-direction along the flow, the $y-$direction across the flow, and the $z$-direction aligned to the gravitational field. To simplify the testing of the new inversion method, the SIA is restricted to the unidirectional case. See Fig. \ref{fig:variables} for a pictorial description of standard SIA notation.

The ice sheet height, $H$, is related to the surface $S$ and the bedrock elevation $b$ via 
\begin{align}
H = S-b
\label{eq:HSzb}
\end{align}
at any time, $t$. 
By considering the momentum balance, volume flux, and mass conservation of the ice sheet, the SIA expression for ice thickness evolution is
\begin{align}
\frac{\partial H}{\partial t} &= f - \nabla \cdot \myvect{q},
\label{eq:dhdt(half)}
\end{align}
where $f$ is the accumulation/ablation function for the ice sheet in meters of water equivalent per year, and
\begin{align}
\myvect{q} &= \int_{b}^{S} \myvect{u} \quad dz
\label{eq:qx(half)}
\end{align}
describes the ice flux by integrating the velocity of the ice along the $x$-direction, $u_x$, from the bedrock to the free surface. Following \cite{Gessese2015} and adapting to include basal slip velocity $u_b$, the velocity profile is given by
\begin{align}
\myvect{u}(z) &= \frac{1}{2} A(\rho g)^3 || \nabla S||_2^2 \nabla S \left[(S-z)^4 - H^4\right] + \myvect{u}_b,
\label{eq:u_x(half)}
\end{align}
where $||\cdot||_2$ is the regular $L2-$norm, $\nabla  S = \left(\frac{\partial S}{\partial x}, \frac{\partial S}{\partial y} \right)$, $\rho$ is the ice density, $g$ is the acceleration due to gravity and $A$ Glens' flow parameter. Values for these constants are given in Table \ref{tab:constants}. The value for $\rho$ is taken as the midpoint of the range for ice sheets as recommended by \citet[][Table 2.1]{Cuffey2010}. The value for A given in Table \ref{tab:constants} is for an ice sheet at $-5 \deg C$ and was recommended by \citet[][Table 3.4]{Cuffey2010}.

The no-slip condition classically imposed \cite[][]{Barcilon1993, Wilchinsky2001, Adhikari2011, Gessese2015, Heining2016} forces $\myvect{u}_b = 0$ for the ice sheet. 
This reduces the amount of surface data required for the inverse problem as without slip the system has only one unknown to recover. However, as discussed in the introduction, basal slip can have significant effect on ice height
which reduces the practical applications if it is neglected.
Here, no such condition is imposed and the ice sheet is allowed to have variable basal slip along the base of the flow.

\citet{Weertman1957} first proposed a power-type law for basal shear on a hard bed and both \citet{Fowler1987} and \citet{Lliboutry1968} proposed a more general form of the law for a flow with cavity formation.
\citet{Budd1979} found this generalised form to be empirically true for ice flow with basal shear described by
\begin{align}
\myvect{\tau}_b^3 = \frac{1}{A_s} \myvect{u}_b,
\label{eq:weertman}
\end{align}
where $\myvect{\tau_b}$ is the basal shear stress, $A_s$ the sliding constant given in Table \ref{tab:constants}, and $\myvect{u_b}$ the basal velocity. The value for $A_s$ is taken from \cite{Gessese2015}.


Pairing this relation with the expression for basal shear from the full derivation of the SIA gives
\begin{align}
\tau_b = -\rho g H \nabla S,
\label{eq:tbderivation}
\end{align}
which combines with eq. \eqref{eq:weertman} to give the following expression for basal velocity
\begin{align}
\myvect{u}_b = - \beta A_s (\rho g)^3 H^3 || \nabla S||_2^2 \nabla S,
\label{eq:ub}
\end{align}
where $\beta(x,y)$  is the basal slip distribution which regulates the amount of basal slip at the ice base. 
Basal slip is restricted such that $\beta(x) \in [0,1]$ for all $x$ in the ice sheet domain. 
Physically, $\beta(x) = 0$ represents a sticky base and $\beta(x) = 1$ a friction-less base. 
It is not required for $\beta(x)$ to be constant along the ice sheets length. 
%

Combining eqns. \eqref{eq:u_x(half)} and \eqref{eq:ub} gives a full expression for the velocity profile. 
This velocity profile is substituted into eq. \eqref{eq:qx(half)} to give the ice flux. Finally, substituting this ice flux into the mass balance gives a non-linear diffusion equation
\begin{align}
\frac{\partial H}{\partial t} &= f - \nabla \left(  D \nabla S \right)
\label{eq:dhdt}
\end{align}
with non-linear effective diffusion coefficient $D$ given by 
\begin{align}
D = \frac{2}{5} A (\rho g)^3  ||\nabla S||_2^2 H^4 \left(H + \frac{5}{2} A_r \beta \right),
\end{align} 
where $A_r$ is the ratio $ \frac{A_s}{A}$. Note that the full velocity profile easily gives an expression for the surface velocity by setting $z = S$:
\begin{align}
\myvect{u_s} &= - \frac{1}{2} A (\rho g)^3 ||\nabla S||_2^2 H^3\left(H + 2 A_r \beta \right) \nabla S.
\label{eq:us}
\end{align}

\begin{table}
	\centering
	\caption{Typical values of constants used throughout.}
	\label{tab:constants}
	\begin{tabular}{lll}
		\hline\noalign{\smallskip}
		Symbol & Name & Value \\
		\noalign{\smallskip}\hline\noalign{\smallskip}
		$A_s$ & Sliding coefficient & 5 $\times 10^{-14} \text{m}^{8} \text{ N}^{-3}\text{ yr}^{-1}$ \\ 
		$A$ & Glen's law parameter & 4.16 $\times 10^{-17} \text{Pa}^{-3} \text{ yr}^{-1}$\\
		$\rho$ & Ice density & 880 kg m$^{-3}$ \\
		$g$ & Gravitational acceleration & 9.81 m s$^{-2}$\\
		\noalign{\smallskip}\hline
	\end{tabular}
\end{table}

\section{Inverse problem methodology}
\label{sec:inverse_methodology}
To begin considering the inverse problem, a methodology is first needed to produce synthetic surface data for a variety of test cases. The approach used for this is outlined in Subsec. \ref{subsec:synth-data}. Once these synthesised surfaces are produced, the inverse methodology can be applied. The approach used for the inverse problem requires two distinct stages; (1) $S \to D$, and (2), $(D, u_s) \to (H, \beta)$, which are described in Subsec. \ref{subsec:lagrangian} and \ref{subsec:newtons} respectively. 

\begin{figure}[ht]
	\centering
	\includegraphics[width=0.7\textwidth]{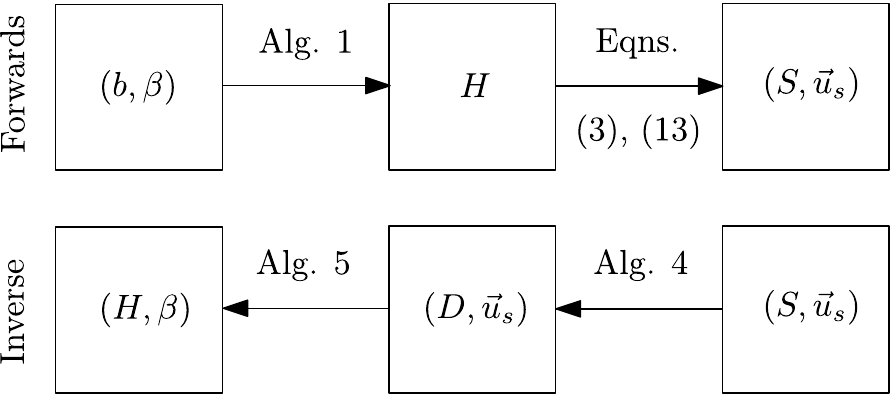}
	\caption{Problem approach with two main phases; (1) forwards, and (2) inverse.}
	\label{fig:problem-stages}
\end{figure}

The  full process described above is outlined in Fig. \ref{fig:problem-stages}. Results of applying this process for each test case as given in the next section, Sec. \ref{sec:numerical_results}.

\subsection{Synthetic data generation}
\label{subsec:synth-data}
To model the ice surface, eq. \eqref{eq:dhdt} needs to solved. There are numerous ways to approach this sort of time dependent diffusion problem and here we use a finite element methodology. First, eq. \eqref{eq:dhdt} is converted into a variational equation following from \citeauthor{Langtangen2016} \cite{Langtangen2016} in Subsec. \ref{subsec:var-form}. The variational equation is solved numerically at each time-step until a steady state ice thickness is reached. This steady state ice thickness is then added to the input bedrock to give a steady state ice surface. Some example surfaces are given in Subsec. \ref{subsec:synth-surfaces-examples}.

\subsubsection{Formulation as a variational problem}
\label{subsec:var-form}
First, a backward Euler discretization is used on the time derivative in eq. \eqref{eq:dhdt} to get
\begin{align}
\frac{H^{n+1} - H^n}{\Delta t} = \nabla (D \nabla S)^{n+1} + f^{n+1}.
\end{align}
Arranging the unknowns to the left gives
\begin{align}
H^{n+1} - \Delta t \nabla(D \nabla S)^{n+1} = H^{n} + \Delta t f^{n+1}.
\end{align}
Next, multiply the above through by a test function $v \in \hat{V}$, where the test space $\hat{V}$ is defined as
\begin{align}
	\label{eq:spaces-pause}
	\hat{V} = \{ v \in H^1(\Omega) \quad : \quad v = 0 \text{ on } \partial \Omega\},
\end{align}
such that the test function, $v$, vanishes on the boundary. The space, $H^{1}(\Omega)$ is the Sobolev space containing functions $v$ such that $v^2$ and $||\nabla v||_2^2$ have finite integrals over the domain $\Omega$. Taking eq. \eqref{eq:spaces-pause} and integrating over the domain gives
\begin{align}
 \int_\Omega H^{n+1} v \dif x - \Delta t \int_\Omega \nabla(D \nabla S)^{n+1} v \dif x = \int_\Omega \left( H^{n}  + \Delta t f^{n+1} \right) v \dif x
\end{align}
where $\dif x$ denotes the differential element for integration over the domain $\Omega$.  Considering only the second order term and applying Green's first identity,
\begin{align}
 - \int_\Omega \nabla \cdot (D \nabla S)v \dif x = \int_\Omega D \nabla S \cdot \nabla v \dif x - \int_{\partial \Omega} D \frac{\partial S}{\partial n} v \dif s
\end{align}
where $\partial \Omega$ is the boundary of $\Omega$, $\frac{\partial S}{\partial n} = $$\myvect{n} \cdot \nabla u$ is the derivative of $S$ in the outward unit normal direction, $\myvect{n}$, and $\dif s$ denotes the differential element for integration over the boundary of $\Omega$. Since the test function, $v$, is required to vanish on the boundary $\partial \Omega$, the second term vanishes also giving
\begin{align}
 \int_\Omega H^{n+1} v \dif x + \Delta t \int_\Omega D^{n+1} \nabla S^{n+1} \cdot \nabla v \dif x = \int_\Omega \left(H^n + \Delta t f^{n+1} \right) v \dif x.
\end{align}
Hence, our final weak form of eq. \eqref{eq:dhdt} is
\begin{align}
\label{eq:weak-form}
F(H;v) &:= \int_\Omega Hv + \Delta t D \nabla S \nabla v - \left(H^n + \Delta t f\right) v \dif x = 0
\end{align}
where all functions are evaluated at the $n+1$ time step unless otherwise stated. By requiring this weak form to hold for all $v \in \hat{V}$, the problem of finding some $H \in V$, the trial-space, is well defined. Hence, the proper problem statement in weak form is: find $H \in V$ such that
\begin{align}
F(H;v) &= 0  \qquad \forall v \in \hat{V}
\end{align}
where
\begin{align}
	\hat{V} &= \{ v \in H^1(\Omega) \quad : \quad v = 0 \text{ on } \partial \Omega\}, \\
	V &= \{ v \in H^1(\Omega) \quad : \quad v =  H_D\text{ on } \partial \Omega\}.
\end{align}
To approximate the solution to this continuous problem, the infinite dimensional spaces $V$ and $\hat{V}$ are replaced with discrete, finite dimensional trial and test spaces, $V_h \subset V$ and $\hat{V}_h \subset \hat{V}$. The discrete problem is then: find $H_h \in V_h \subset V$ such that
\begin{align}
F(H_h;v_h) &= 0  \qquad \forall v_h \in \hat{V}_h \subset \hat{V}.
\end{align}
This variational problem, together with suitable choices of function spaces $V_h$ and $\hat{V}_h$, uniquely defines the approximate solution $H_h$ to eq. \eqref{eq:dhdt}.

\subsubsection{Numerical computations and test case classification}
\label{subsec:synth-surfaces-examples}
A simple time stepping iteration is implemented to compute the steady state ice profile as described in Alg. \ref{alg:forwards}. The weak problem, eq. \eqref{eq:weak-form}, is solved using the open source finite element computational software libary, FEniCS \cite{LoggMardalEtAl2012a, AlnaesLoggEtAl2012a}. FEniCS provides a large libary of finite elements and numerical solvers. In this study, $P1$ elements are used for both spaces and the systems are solved using a GMRES linear solver, which is part of the PETSc package \cite{petsc-web-page, petsc-user-ref, petsc-efficient}.

For all test cases, the accumulation/ablation function, $f$ is defined as
\begin{equation}
f(x) = \begin{cases} f_0 \left (1- \frac{300 - x}{100} \right) \quad \text{if} \quad x \leq 300 \\ f_0 \left (\frac{2200 - x}{1900} \right) \quad \text{if} \quad x \geq 300 \end{cases}
\end{equation}
where $f_0$ is the maximum value of the accumulation/ablation function and set to 0.5 for all future calculations. Adjusting this maximum values simply raises or lowers the steady state surface \cite{LeMeur2004}.
This function gives the most accumulation at the top end of the glacier and then linearly decreases along it's length until at the bottom end which has net ablation.

The bedrock and basal slip profiles are each chosen from three classes and there are three cases for each class. Equations which describe these profiles are given below in the lists following. For each class, the parameter changed to give a new case is $\gamma$. This may change the slope, extent or height dependent on the equation.

The bedrock profiles are given by;
\begin{enumerate}
	\item Inclined. Denoted by $b_1$, and defined by
	\begin{align*}
		b(x) = (4500 \gamma) - \gamma x,
	\end{align*}
	where changing $\gamma$ changes the slope and taken from  $\gamma \in \{0.15, 0.2, 0.25 \}$.
	\item Bump. Denoted by $b_2$, and defined by
	\begin{align*}
		b(x) \!\begin{aligned}[t]
		&=900 - 0.2x + \gamma \left(50e^{\frac{-(x-2000)^2}{300^2}} \right)
		\end{aligned}
	\end{align*}
	where changing $\gamma$ affects the bump height, $\gamma \in \{1,2,3\}$.
	\item Undulations. Denoted by $b_3$, and defined by
	\begin{align*}
		b(x) \!\begin{aligned}[t]
		&=900 - 0.2x + \gamma \left(-40e^{\frac{-(x-1300)^2}{300^2}} + 60e^{\frac{-(x-3100)^2}{400^2}}\right)
		\end{aligned}
	\end{align*}
	where $\gamma$ again dictates bump height and $\gamma \in \{1,2,3\}$.
\end{enumerate}

The basal slip profiles are given by;
\begin{enumerate}
	\item Constant. Denoted by $\beta_1$, and defined by
	\begin{align*}
	\beta(x) = \gamma,
	\end{align*}
	with$\gamma \in \{0, 0.5, 1\}$.
	\item Gaussian. Denoted by $\beta_2$, and defined by
	\begin{align*}
	\beta(x) = e^{-\left(\frac{x-2500}{\gamma}\right)^{10}}
	\end{align*}
	where changing $\gamma$ affects the bump extent, $\gamma \in \{500,1000,1500\}$.
	\item Switch. Denoted by $\beta_3$, and defined by
	\begin{align*}
	\beta(x) = \left(\frac{1}{2} + \frac{1}{2} \text{erf}\left(\frac{x-2500}{\gamma} \right) \right)
	\end{align*}
	where $\gamma$ again changes the extent and $\gamma \in \{500,1000,1500\}$.
\end{enumerate}

\begin{figure}[ht]
	\centering
	\includegraphics[width=\textwidth]{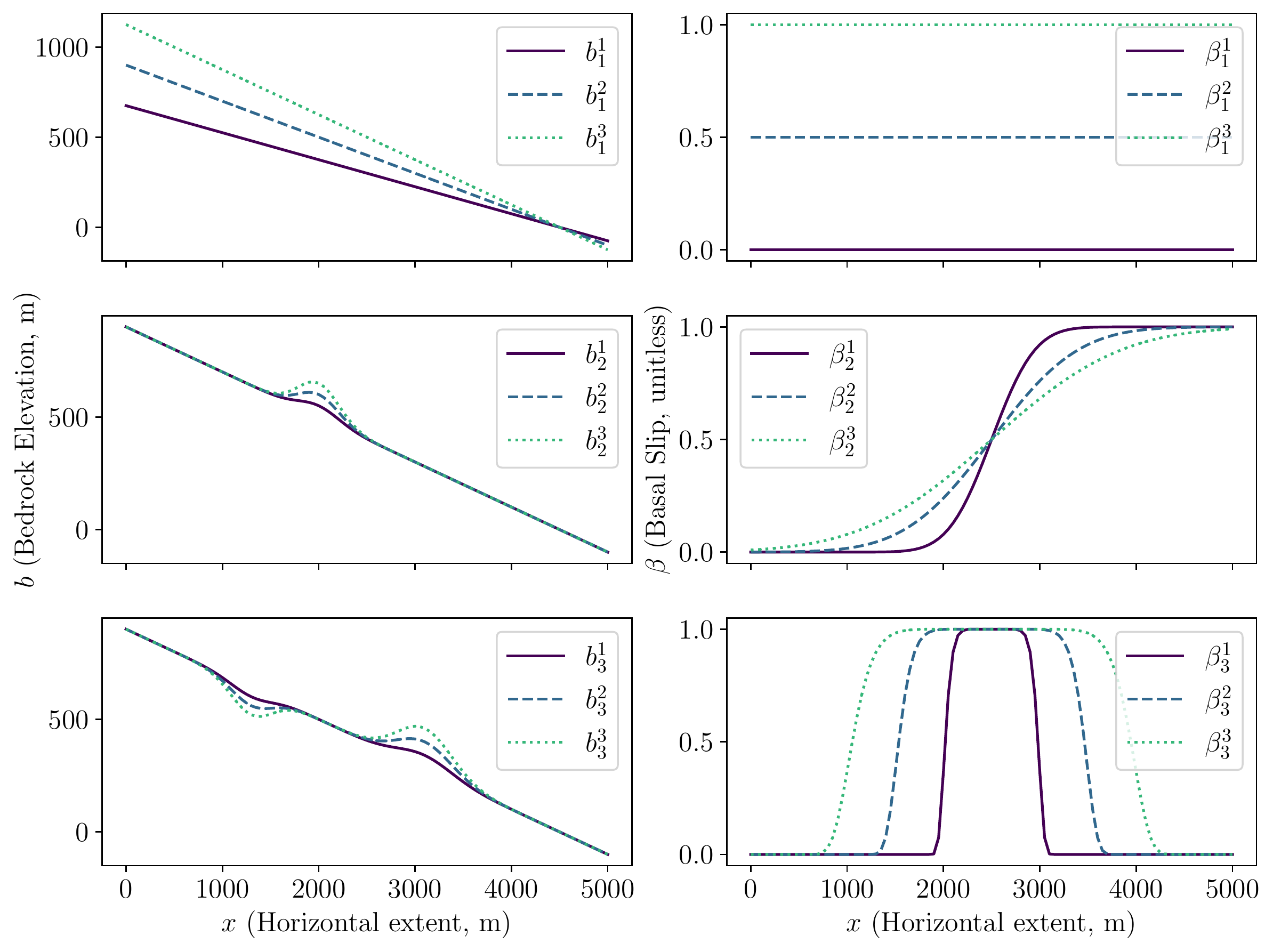}
	\caption{Graphical representation of the different bedrock and slip profiles which give rise to separate synthetic test cases.}
	\label{fig:case-visuals_input}
\end{figure}

Going forward, the particular combination of bedrock and basal slip profiles paired to produce a case of synthetic test data will be denoted by $(b, \beta)$ with subscripts giving the class of profile and the superscripts the particular case (by selection of $\gamma$). Visualisation of each profile is given in Fig. \ref{fig:case-visuals_input}.

\begin{figure}[ht]
	\centering
	\includegraphics[width=0.7\textwidth]{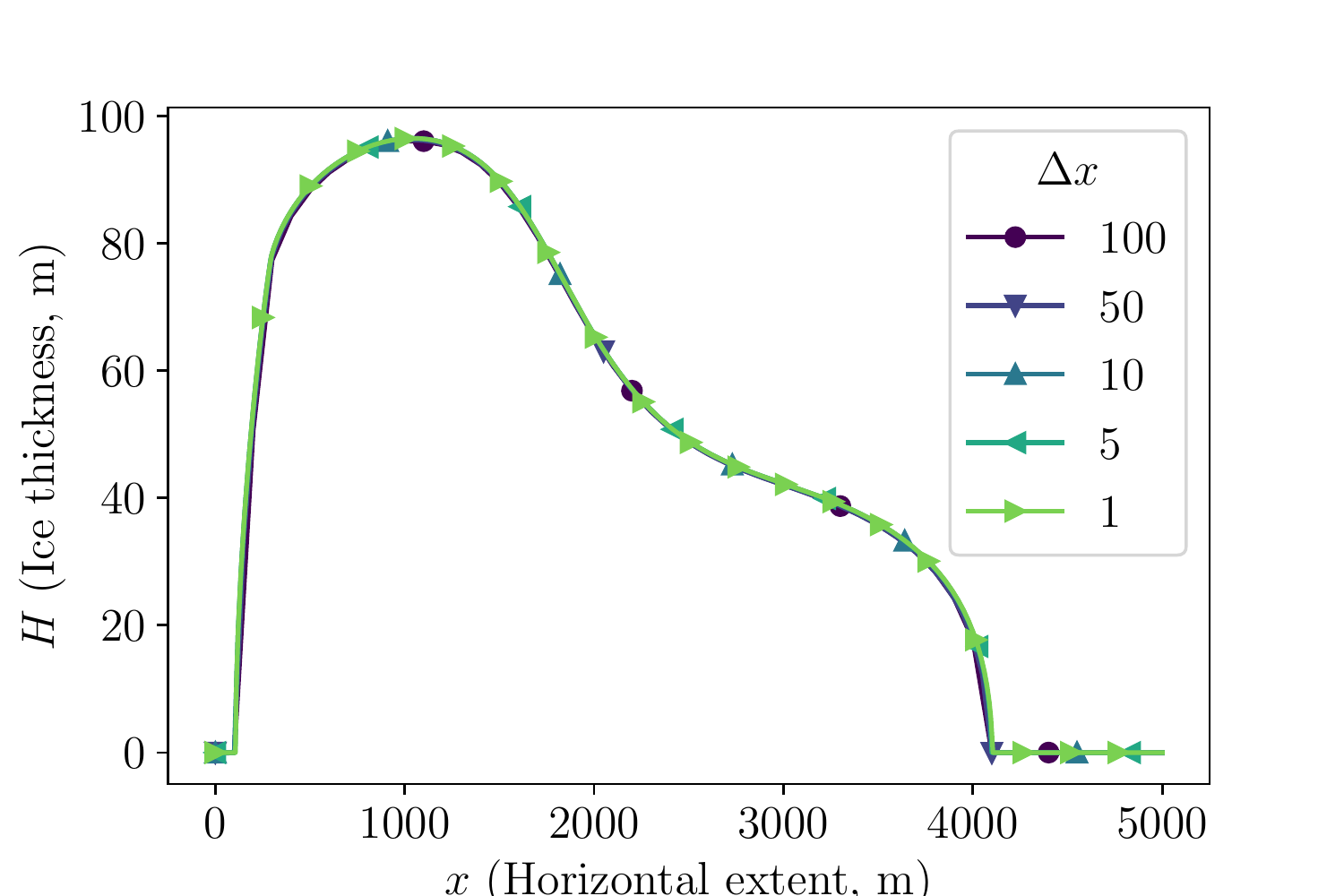}
	\caption{Steady state ice thickness in the forwards problem for $b_1^1$ together with $\beta_2^1$ for different values of $\Delta x$ as indicated in the legend.}
	\label{fig:mesh-independence}
\end{figure}

Modelled steady state ice thickness profiles for given bedrock and slip profiles match the previous literature \cite{LeMeur2004, Gessesse2014, Gessese2015, McGeorge2021} which were computed using finite difference schemes. The scheme is mesh independent as is clear in Fig. \ref{fig:mesh-independence}. Henceforth, all steady state profiles plotted and used are computed on a mesh with $\Delta x = 20$ and $\Delta t = 10^{-2}$. A selection of steady state profiles is given below in Fig. \ref{fig:case-visuals-output} to illustrate effects observed at the surface due to changing basal conditions. 

\begin{algorithm}
	\caption{\textsc{Forwards SIA}}
	\label{alg:forwards}
	\vspace{1em}
	\begin{enumerate}[label = {\textbf{\arabic*}.}]
		\item Set $H^{0} = 0$ and $n=0$.
		\item Compute $H$ such that 
		\begin{align*}
		F(H;v) &= \int_\Omega Hv + \Delta t D(H) \nabla H \nabla v - \left[H^n + \Delta t f\right] v \dif x = 0
		\end{align*}
		
		To ensure $H \geq 0$, set $H^{n+1} = \max\{0, H\}$.
		
		If $H_\Delta = \frac{H^{n+1} - H^n}{\Delta t} < 10^{-1}$, set $H^* = H^{n+1}$ and \textsc{goto} 3.
		
		Otherwise, set $n = n+1$, \textsc{repeat} 2.
		\item Finally compute $S^*$ and $\boldsymbol{u_s}^*$ via
		\begin{align*}
		S^* &= H^* + b, \\
		\boldsymbol{u_s}^* &= - \frac{1}{2} A (\rho g)^3 ||\boldsymbol{\nabla S}^*||_2^2 {H^*}^3\left(H^* + 2 \frac{A_s}{A} \beta \right) \boldsymbol{\nabla S}^*.
		\end{align*}
	\end{enumerate}
\end{algorithm}

\begin{figure}[ht]
	\centering
	\includegraphics[width=\textwidth]{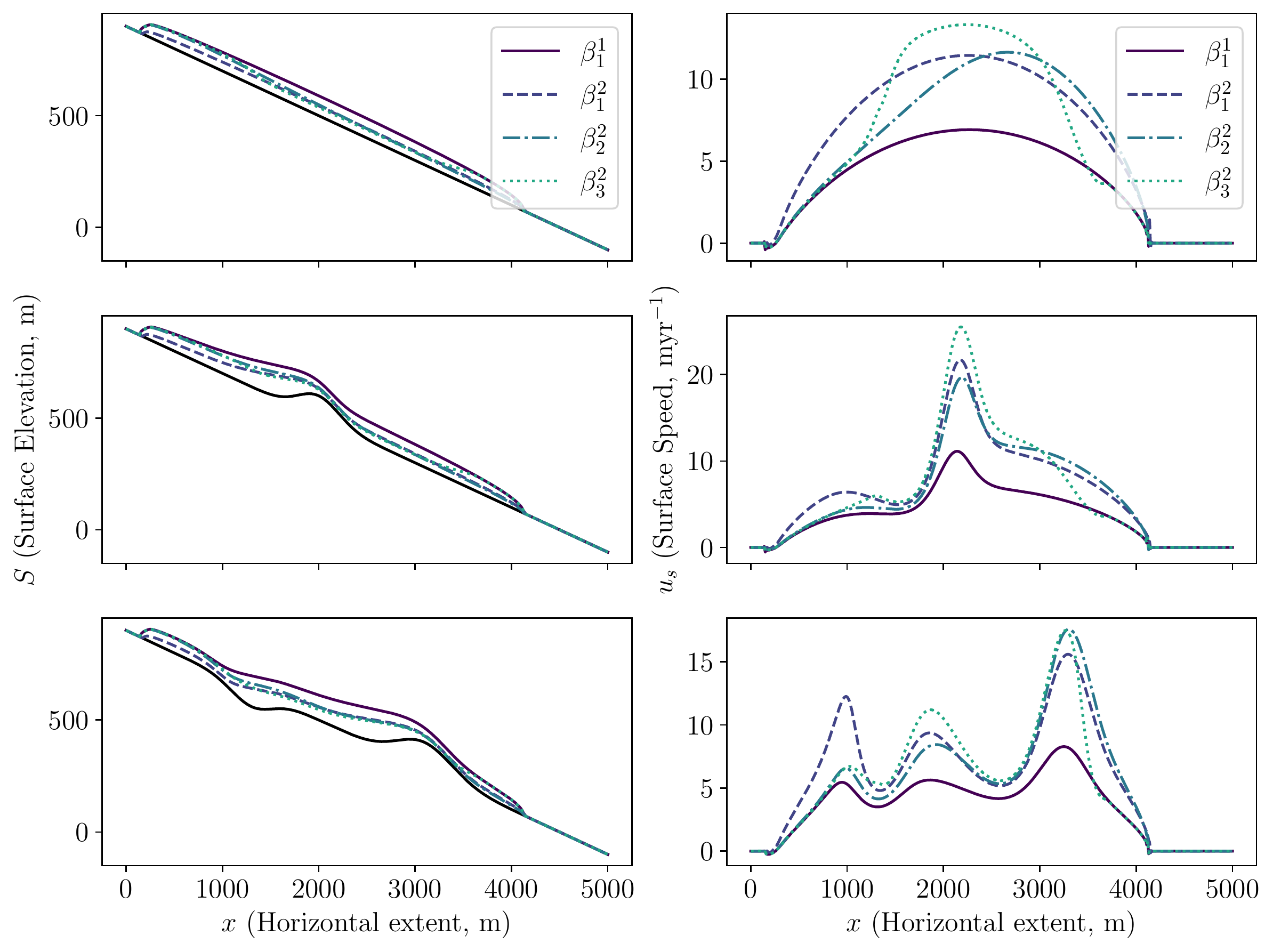}
	\caption{Left: resultant steady state ice surfaces for different slip regimes (indicated by line-style/colour in the plot legend) paired together with different bedrock profiles (indicated in black beneath the plotted surfaces) . Right: corresponding surface speed, $u_s$ for each steady state surface..}
	\label{fig:case-visuals-output}
\end{figure}

\begin{figure}[ht]
	\centering
	\includegraphics[width=0.5\textwidth]{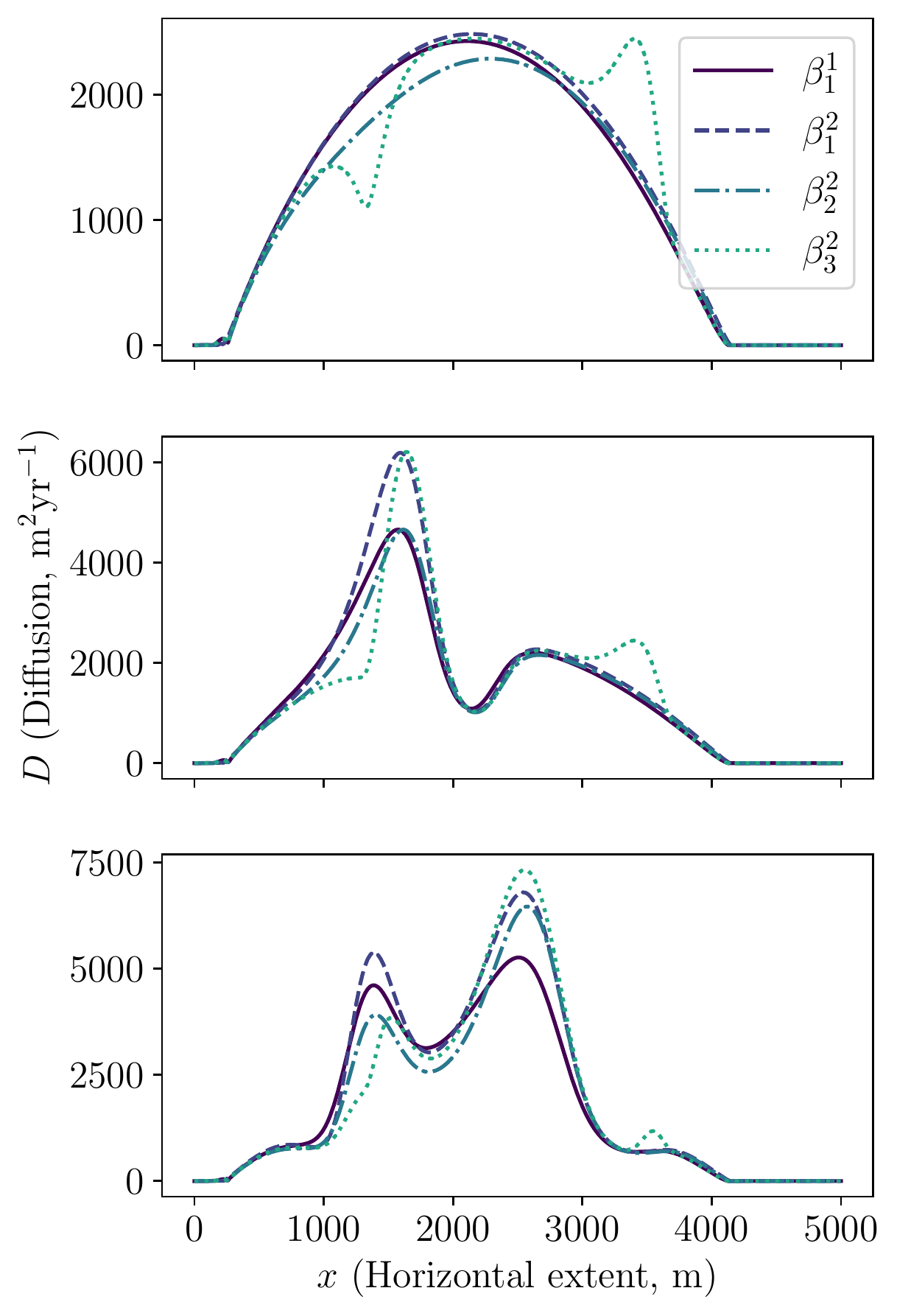}
	\caption{Effect of different slip regimes on the steady state diffusion.}
	\label{fig:steady-state-D-target}
\end{figure}

\subsection{Recovery of $D$ from surface data}
\label{subsec:lagrangian}
To recover the diffusion coefficient, $D$, from data an optimisation approach is used. This approach to minimise the error between observations of the surface and the one modelled using the recovered $D$. Hence the optimal control problem is to minimise the objective functional
\begin{align}
\mathcal{J}(S, D) &= \mathcal{J}_{mis} + \mathcal{J}_{reg} \\
&= \frac{1}{2}\int_\Omega ||S_{\text{obs}} - S ||_2^2 \dif \Omega + \alpha \int_\Omega ||\nabla D||_2^2 \dif \Omega
\end{align}
with respect to $D$ subject to
\begin{align}
- \nabla ( D \nabla S) &= f \label{eq:SIA-SS}\\
S &= S_{\text{obs}} \text{ on } \Gamma\\
\Omega &= [x_s, x_f]\\
\Gamma &= (x_s) \cup (x_f)
\end{align}
\vspace{1ex}
where $x_s$ and $x_f$ demarcate the start and end of the ice domain. The Tikhonov regularisation term, $\mathcal{J}_{reg}$, is necessary to ensure that the problem is well-posed for poor initial conditions or data. The Tikhonov term can be thought of as a cost term for the gradient of the control; essentially the larger $\alpha$, the more favourable a smooth solution is.

The PDE constraint $- \nabla ( D \nabla S) = f$ can equivalently be considered as a residual constraint $e(D, S) = 0$ where $e(D, S)$ is defined by the residual equation (in the weak sense):
\begin{align}
\label{eq:e-equation}
(\nabla e(D, S), \nabla \phi) = (D \nabla S, \nabla \phi) - (f, \phi) \quad \forall (D, S) \in K \times V, \quad \phi \in V
\end{align}
where
\begin{align*}
V &= H^1(\Omega), \\
K &= \left\{ D \in L^1(\Omega); \quad \int_\Omega ||\nabla D||_{2} \dif \Omega < \infty \text{ and } \alpha_1 \leq D \leq \alpha_2 \text{ a.e. in } \Omega \right\}.
\end{align*}

To minimise the objective, an augmented Lagrangian approach is used. This enables relaxation of the residual constraint $e(D, S)$ which enhances the convexity of the objective functional. To do this, the augmented Lagrangian functional, $\mathcal{L}_r: K \times V \times V \to R$, is introduced:
\begin{align}
\mathcal{L}_r (D, S; \mu) = \mathcal{J}(D, S) + (\nabla \mu, \nabla e (D, S)) + \frac{r}{2}||\nabla e(D, S) ||_{2}^2
\end{align}
where $r \geq 0$ is a given constant and $R$ denotes the reals. Finding the saddle point of this augmented form is equivalent to finding a minimum of the objective functional \cite[e.g.][]{Keung2000, Chen1999}.

Following from \cite{Keung2000}, the augmented Lagrangian is discretised. Taking $V_h$ as the standard piecewise linear finite element space, the discrete form , $L_r: K_h \times \mathring{V_h} \times \mathring{V_h} \to R$, is given by
\begin{align}
L_r(D_h, S_h; \mu_h) &= J_h(D_h, S_h) + (\nabla \mu_h, \nabla e_h (D_h, S_h)) + \frac{r}{2}||\nabla e_h(D_h, S_h) ||_{2}^2, 
\end{align}
where
\begin{align}
D_h \in K_h &= K \cap V_h, \quad S_h \in \mathring{V_h} = V_h \cap H^1(\Omega),
\end{align}
with
\begin{align}
J_h(D_h, S_h) &= \frac{1}{2}\int_\Omega ||S^{\text{obs}} - S_h ||_2^2 \dif \Omega + \alpha \int_\Omega ||\nabla D_h||_2^2 \dif \Omega.
\end{align}
Taking the discrete form of the residual equation \eqref{eq:e-equation}, $e_h(D_h, S_h) \in \mathring{V_h}$ is defined as the solution to
\begin{align}
	(\nabla e_h(D_h, S_h), \nabla \phi) &= (D_h \nabla S_h, \nabla \phi_h) - (f, \phi_h), \quad \forall \phi_h \in \mathring{V_h}
\end{align}
for any $(D_h, S_h) \in K_h \times \mathring{V_h}$. It can be shown (see e.g. \cite{Chen1999}) that for any $r \geq 0$, there exists at least one saddle point of $L_r$ which can be found using a simple Uzawa type algorithm given in Alg. \ref{alg:basic-uzawa}. This basic form of the algorithm is convergent for choices of $0<\rho_r<r$ \parencite{Chen1999}. 

\begin{algorithm}
	\caption{\textsc{Basic Uzawa}}
	\label{alg:basic-uzawa}
	\vspace{1em}
	Choose $r \geq 0$ and $\rho_r < r$.
	Given $\lambda_0 \in \mathring{V}_h$, iterate through $n$ by computing the pair $\{D_h, S_h\} \in $ such that
	\begin{align}
	\label{eq:uzawa step 2}
	L_r(D^n, S^n; \lambda^n) = \min\{L_r(p, v; \lambda^n) \quad \forall (p, v) \in K_h \times \mathring{V}_h \}
	\end{align}
	and then updating $\lambda^{n+1}$ via
	\begin{align}
	\lambda^{n+1} = \lambda^n + \rho_r e_h(D^n, S^n).
	\end{align}
\end{algorithm}

To perform the minimisation in Alg. \ref{alg:basic-uzawa} (eq. \eqref{eq:uzawa step 2}), an alternative iteration is used, first computing $S_h$ and then the corresponding $D_h$. As in \cite{Keung2000}, this is referred to as the modified Uzawa algorithm and is given explicitly in Alg. \ref{alg:mod-uzawa}. Step 2 of this modified form still requires two minimisations steps. The following will show that each minimisation (eq. \eqref{eq: uzawa step 2a} and eq. \eqref{eq: uzawa step 2b}) is equivalent to solving a system of variational equations, which FEniCS is capable of solving.

\begin{algorithm}
	\caption{\textsc{Modified Uzawa}}
	\label{alg:mod-uzawa}
	\vspace{1em}
	Choose $r \geq 0$ and $\rho_r < r$.
	Given $\lambda_0 \in \mathring{V}_h$ and $D^0 \in K_h$. Set $n=1$.
	\begin{enumerate}[label = {\textbf{\arabic*}.}]
		\item Set $k=1$ and $D_0^n = D^{n-1}$ .
		\item Compute $S_{k}^{n} \in \mathring{V}_h$ by solving
		\begin{align}
		\label{eq: uzawa step 2a}
		L_r(D_{k-1}^{n}, S_{k}^{n}; \lambda^{n-1}) = \min_{v_h \in \mathring{V}_h} L_r(D_{k-1}^{n}, v_h; \lambda^{n-1})
		\end{align}
		and then compute $D_{n}^{k} \in V_h$ by solving
		\begin{align}
		\label{eq: uzawa step 2b}
		L_r(D_{k}^{n}, S_{k}^{n}; \lambda^{n-1}) = \min_{p_h \in {V}_h} L_r(p_h, S_{k}^{n}; \lambda^{n-1})
		\end{align}
		Compute $D_k^n = \max\{\alpha_1, \min\{D_k^n, \alpha_2\}\}$.
		
		If $||D_k^n - D_{k-1}^{n}||_2 \leq \epsilon_q$ \textsc{or} $k \geq k_{\max}$,  set $S^n = S_k^n$ and $D^n = D_k^n$, \textsc{Goto} 3.
		
		Otherwise, set $k=k+1$, \textsc{Goto} 2.
		\item Compute $\lambda^n$ by
		\begin{align}
		\lambda^n = \lambda^{n-1} + \rho_r e_h(D^n, S^n).
		\end{align}
		If $||S^n - S^{\text{obs}}||_2 \leq \epsilon_S$ \textsc{or} $n \geq n_{\max}$, \textsc{End}.
		
		Otherwise, set $n= n+1$, \textsc{Goto} 1.
	\end{enumerate}
\end{algorithm}

The Gateaux derivative of $\mathcal{L}_r$ with respect to  $S_h$ in the direction $w_h$ is given by
\begin{align}
\label{L'-u(L2)}
\mathcal{L}_r'(D_h, S_h; \lambda_h) w_h &= (S_h-S^{\text{obs}}, w_h) + (D_h \nabla \lambda_h, \nabla w_h) + r (D_h \nabla e_h, \nabla w_h).
\end{align}
To minimise $\mathcal{L}_r$, we solve the above two equations as a system by setting eq. \eqref{L'-u(L2)} to 0:
\begin{align}
0 & = \mathcal{L}_r'(D_h, S_h; \lambda_h) w_h, \\
0 &= (S_h-S^{\text{obs}}, w_h) + (D_h \nabla \lambda_h, \nabla w_h) + r (D_h \nabla e_h, \nabla w_h). \\
\end{align}
Recall that $e_h(D_h, S_h)$ is the solution to
\begin{align*}
(\nabla e_h(D_h, S_h), \nabla \phi) &= (D_h \nabla S_h, \nabla \phi) - (f, \phi).
\end{align*}
Hence, arranging the known variables to the left (terms not involving either of $e_h$ or  $S_h$) our minimisation is equivalent to finding $(S_h, e_h) \in \mathring{V}_h \times \mathring{V}_h$ such that
\begin{align}
(S_h, w_h) + r(D_h, \nabla e_h, \nabla w_h) &= (z, w_h) - (D_h \nabla \lambda_h, \nabla w_h) \\
(\nabla e_h(D_h, S_h), \nabla \phi_h) &= (D_h \nabla v_h, \nabla \phi_h) - (f, \phi_h),
\end{align}
$\forall w_h \in \mathring{V_h}$ and $\forall \phi_h \in \mathring{V_h}$. 

Similarly, the Gateaux derivative of $\mathcal{L}_r$ with respect to $D_h$ in the direction $p_h$ is given by
\begin{align}
\label{eq:L'-q(L2)}
\mathcal{L}_r'(D_h, S_h; \lambda_h) p_h &= \alpha (\nabla D_h, \nabla p_h) + (p_h \nabla S_h, \nabla \lambda_h) + r (p_h \nabla S_h, \nabla e_h)
\end{align}
where $e_h$ is the solution to eq. \eqref{eq:e-equation} as above. As above, setting eq. \eqref{eq:L'-q(L2)} to 0 and arranging unknowns to the left gives the system which can be solved to find $D_h$ which minimises $\mathcal{L}_r$ with respect to $D_h$. The problem is then to find $(D_h, e_h) \in V_h \times \mathring{V}_h$ such that
\begin{align}
\alpha (\nabla D_h, \nabla p_h) + r (p_h \nabla S_h, \nabla e_h) &= -(p_h \nabla S_h, \nabla \lambda_h), \\
(\nabla e_h, \nabla \phi_h) - (D_h \nabla v_h, \nabla \phi) &=  - (f, \phi_h),
\end{align}
$\forall p_h \in V_h$ and $\forall \phi_h \in \mathring{V_h}$.
Hence, the two minimisation problems in Step 2 of Alg. \ref{alg:mod-uzawa} can be expressed as solving two systems as is outlined in Alg. \ref{alg:mod-uzawa-sys}.

\begin{algorithm}
	\caption{\textsc{Modified Uzawa (System Version)}}
	\label{alg:mod-uzawa-sys}
	\vspace{1em}
	Choose $r \geq 0$ and $\rho_r < r$.
	Given $\lambda_0 \in \mathring{V}_h$ and $D^0 \in K_h$. Set $n=1$.
	\begin{enumerate}[label = {\textbf{\arabic*}.}]
		\item Set $k=1$ and $D_0^n = D^{n-1}$ .
		\item Compute the pair $(S_{n}^{k}, e_h) \in \mathring{V}_h \times \mathring{V}_h$ by solving
		\begin{align}
		\label{eq: uzawa step 2a sys}
		(S_{k}^{n}, w_h) + r(D_{k-1}^{n}, \nabla e_h, \nabla w_h) &= (S^{\text{obs}}, w_h) - (D_{k-1}^{n} \nabla \lambda^{n-1}, \nabla w_h) \\
		(\nabla e_h, \nabla \phi) &= (D_{k-1}^{n} \nabla v_h, \nabla \phi) - (f, \phi),
		\end{align}
		and then compute $(D_{n}^{k}, e_h) \in V_h \times \mathring{V}_h$ by solving
		\begin{align}
		\label{eq: uzawa step 2b sys}
		\alpha (\nabla D_{k}^{n}, \nabla p_h) + r (p_h \nabla S_{k}^{n}, \nabla e_h) &= -(p_h \nabla S_{k}^{n}, \nabla \lambda^{n-1}), \\
		(\nabla e_h, \nabla \phi) - (D_{k}^{n} \nabla v_h, \nabla \phi) &=  - (f, \phi).
		\end{align}
		Compute $D_k^n = \max\{\alpha_1, \min\{D_n^k, \alpha_2\}\}$.
		
		If $||D_n^k - D_n^{k-1}||_2 \leq \epsilon_q$ OR $k \geq k_{\max}$,  set $S^n = S_k^n$ and $q^n = D_k^n$, GOTO 3.
		
		Otherwise, set $k=k+1$, GOTO 2.
		\item Compute $\lambda^n$ by
		\begin{align}
		\lambda^n = \lambda^{n-1} + \rho_r e_h(q^n, S^n)
		\end{align}
		where $e_h(D^n, S^n)$ solves
		\begin{align}
		(\nabla e_h(D^n, S^n), \nabla \phi) &= (D^n \nabla S^n, \nabla \phi) - (f, \phi).
		\end{align}
		If $||D^n - D^{n-1}||_2 \leq 10^{-3}$ OR $n \geq n_{\max}$, END.
		
		Otherwise, set $n= n+1$, GOTO 1.
	\end{enumerate}
\end{algorithm}

\subsection{Subsequent recovery of $(H, \beta)$ from $D$}
\label{subsec:newtons}

Once $D$ is recovered via the Uzawa algorithm above, one more step is required to recover the ice thickness and basal slip. This is to find $(H, \beta)$ given $(D^{\text{inv}}, u_{s}^{\text{obs}})$. To do this, the SIA expressions for $D$ and $\boldsymbol{u_s}$ are needed. Recall that these were
\begin{align}
0 &= D - K ||\nabla S||_2^2 H^4 (H + \frac{5}{2} A_r \beta) \\
0 &= \myvect{u_s} - \left[\frac{-5}{4} K ||\nabla S||_2^2 H^3 (H + 2 A_r \beta)\right] \nabla S
\end{align}
where $K = 2/5 A (\rho g)^3$, $A_r = A_s / A$, $||\nabla S||_2^2 = \left(\frac{\partial S}{\partial x}\right)^2 + \left(\frac{\partial S}{\partial y}\right)^2$. Since the velocity has two-components, this is a system of three equations with two unknowns. To reduce this to a system of two equations with two unknowns, the two-norm of the velocity is taken. The velocity equation is therefore,
\begin{align}
0 &= ||\myvect{u_s}||_2 - \left[\frac{5}{4} K ||\nabla S||_2^2 H^3 (H + 2 A_r \beta)\right] ||\nabla S||_2
\end{align}
which can be rearranged to give
\begin{align}
\beta &= \frac{1}{2 A_r}\left(\frac{||\myvect{u_s}||_2}{\frac{5}{4} K ||\nabla S||_2^2 H^3 ||\nabla S||_2}  - H \right).
\end{align}
Now, substituting this into the equation involving $D$ gives
\begin{align}
0 &= D - K ||\nabla S||_2^2 H^4 \left(H + \frac{5}{2} A_r \left( \frac{1}{2 A_r}\left(\frac{||\myvect{u_s}||_2}{\frac{5}{4} K ||\nabla S||_2^2 H^3 ||\nabla S||_2}  - H \right) \right) \right) \\
0  &= D - \frac{||\myvect{u_s}||_2}{ ||\nabla S||_2 } H  + \frac{1}{4} K ||\nabla S||_2^2 H^5 
\end{align}
Using this final form, it is clear that recovering $H$ is as simple as solving for the zeroes of the quintic
\begin{align}
p(H) &= \frac{1}{4} K ||\nabla S||_2^2 H^5 - \frac{||\myvect{u_s}||_2}{ ||\nabla S||_2 } H + D.
\end{align}

To improve the efficiency of the  numerical algorithm, the extrema of this polynomial are used. To find these, differentiate and set to zero,
\begin{align}
p'(H_{ex}) =0 &= \frac{5}{4} K ||\nabla S||_2^2 H_{ex}^4 - \frac{||\myvect{u_s}||_2}{ ||\nabla S||_2 }, \\
\implies H_{ex}^4 & = \frac{\frac{||\myvect{u_s}||_2}{ ||\nabla S||_2 }}{ \frac{5}{4} K ||\nabla S||_2^2}, \\
\implies H_{ex} &= \left(\frac{||\myvect{u_s}||_2}{\frac{5}{4} K ||\nabla S||_2^2 ||\nabla S||_2} \right)^{1/4}.
\end{align}
Note that $H_{ex} \geq 0$ always. This extremum can be classified as a minimum or maximum by finding the sign of the second derivative,
\begin{align}
p''(H) &= \frac{20}{4} K ||\nabla S||_2^2 H^3.
\end{align}
Since $H_{ex} \geq 0$ it follows that $p''(H_{ex}) \geq 0$ so the extremum is a local minimum.

Consider again $\beta$ defined by the equation for $\myvect{u_s}$. Since $\beta \geq 0$ it follows that
\begin{align}
0 &\leq \frac{||\myvect{u_s}||_2 - H ( \frac{5}{4} K ||\nabla S||_2^2  ||\nabla S||_2 H^3)}{\frac{10}{4} A_r K ||\nabla S||_2^2  ||\nabla S||_2 H^3} \\
\implies 0 &\leq||\myvect{u_s}||_2 - H ( \frac{5}{4} K ||\nabla S||_2^2  ||\nabla S||_2 H^3)\\
\implies H^4 &\leq\frac{||\myvect{u_s}||_2}{\frac{5}{4} K ||\nabla S||_2^2  ||\nabla S||_2} \\
\implies H &\leq \left(\frac{||\myvect{u_s}||_2}{\frac{5}{4} K ||\nabla S||_2^2  ||\nabla S||_2} \right)^{1/4}.
\end{align}
In fact, when $\beta = 0$, precisely the abscissa of the minimum of $p(H)$ is returned. Hence, this abscissa is the maximum allowable value which $H$ can take to give a viable solution. This notion can be used to restrict the search interval for roots of $p(H)$ to $H \leq H_{ex}$. Additionally, if $p(H_{ex}) \geq 0$, there are no other viable roots (there is one negative root which is not allowed since $H \geq 0$) and so $H^* = H_{ex}$. These processes are outlined in the modified Newton's algorithm given in Alg. \ref{alg:mod-newtons}.

\begin{algorithm}
	\caption{\textsc{Modified Newtons}}
	\label{alg:mod-newtons}
	\vspace{1em}
	For each coordinate $\boldsymbol{x}$ in $\Omega$:
	
	Set $j=0$.
	\begin{enumerate}[label=\textbf{\arabic*}.]
		\item Calculate $H_{\max}$ by
		\begin{align}
		H_{max} = \left(\frac{||\myvect{u_s}||_2}{\frac{5}{4} K ||\nabla S||_2^2  ||\nabla S||_2} \right)^{1/4}
		\end{align}
		if $p(H_{\max}) \geq 0$, set $h^* = H_{\max}$. \textsc{End}.\\
		Otherwise, \textsc{Goto} 2.
		\item Set $n = 0$ and $h^0 = H_{\max} - 5j$. 
		\begin{enumerate}[label = {(\textbf{\alph*})}]
			\item Set $h^n = h^0$.
			\item Compute $h^{n+1}$ by
			\begin{align}
			h^{n+1} = h^n - \frac{p(h^n)}{p'(h^n)}
			\end{align}
			\item If $p(h^{n+1}) \leq \epsilon_p$, set $h_{\text{newt}} = h^{n+1}$. \textsc{Goto} (3).\\
			Otherwise, increment $n= n+ 1$. \textsc{Goto} (a).
		\end{enumerate}
		\item If $h_{\text{newt}} < 0 $ or $h_{\text{newt}} > H_{\max}$, set $j = j+1$. \textsc{Goto} (2). \\
		Otherwise, set $h^* = h_{\text{newt}}$. \textsc{End}.
	\end{enumerate}
\end{algorithm}

\section{Numerical results}
\label{sec:numerical_results}
The results of this study are presented in two parts. The first relating to the recovery of the non-linear diffusion $D$ in Subsec. \ref{subsec:D-recovery}, and the second covering the subsequent recovery of the pair $(H^{\text{inv}}, \beta^{\text{inv}})$ from the pair $(D^{\text{inv}}, u_s^{\text{obs}})$ in Subsec. \ref{subsec:h,beta-recover}. For each recovery, the errors are calculated by
\begin{align}
\label{eq:err_calc}
\mathcal{E}_F = \frac{||F^* - F^{\text{inv}}||_2}{||F^*||_2}, \qquad F \in \{D, H, \beta\}.
\end{align}
In the case where $\beta \equiv \beta_1^1 = 0$, the error is calculated as $\mathcal{E} = ||\beta^{\text{inv}}||_2$.
\subsection{Recovery of $D$}
\label{subsec:D-recovery}
In each implementation, the domain of the problem is $\Omega = [x_s, x_f]$ where $x_s$ is the location of the dome (first point after the onset of ice where $\frac{\partial S}{\partial x} = 0$) and $x_f$ is where the ice ends. The interval is split into uniformly distributed intervals of length $1/{nx}$. Unless otherwise specified, $nx = 200$. The augmented Lagrangian coefficient is set to $r=1$ and the initial guess for the Lagrange multiplier is $\lambda^0 = 0$. The lower and upper bounds used to constrain $K$ are $\alpha_1 = 10^{-2}$ and $\alpha_2 = 10^{5}$. 

\begin{figure}[ht]
	\centering
	\includegraphics[width=0.5\textwidth]{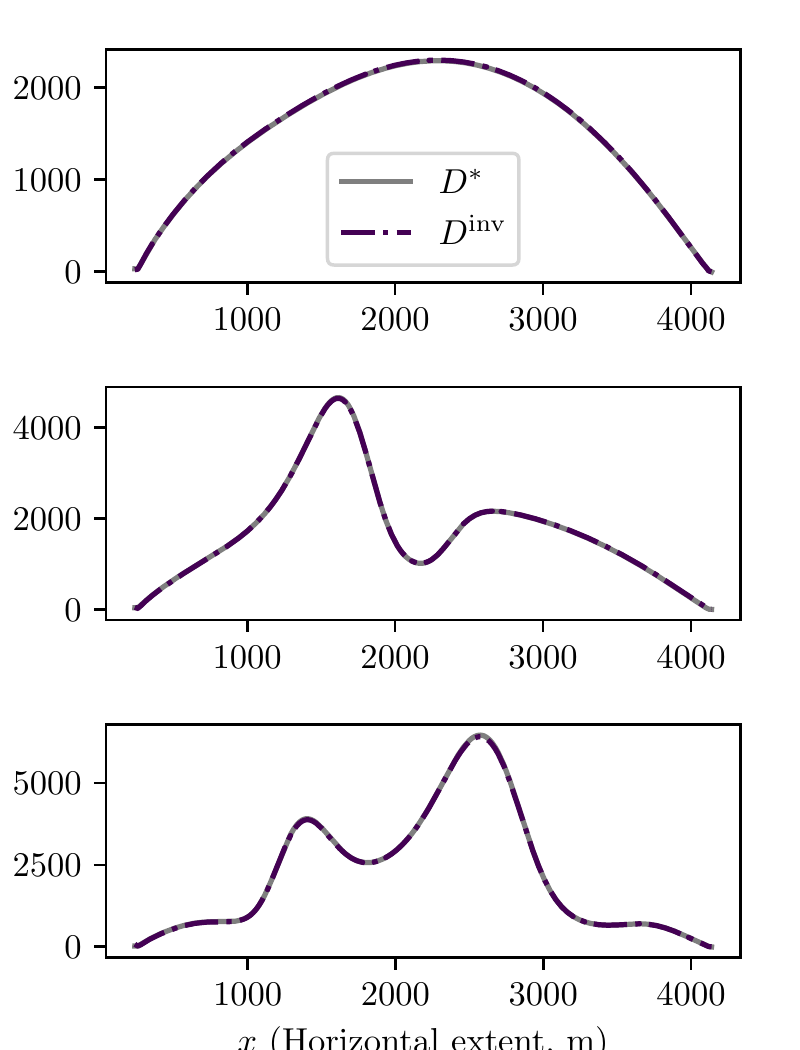}
	\caption{A selection of optimised $D^{\text{inv}}$ from the algorithm. In all cases, the target bedrock is defined by $b \equiv b_i^2$ and $\beta \equiv \beta_2^2$.}
	\label{fig:d-inv-examples}
\end{figure}

To solve the two systems of variational equations in Alg. \ref{alg:mod-uzawa-sys}, FEniCS is used again. For each test case, an initial guess of $D^0 = 1000$ is paired with an initial regularisation $\alpha = 10^0$. Using these inputs, Alg. \ref{alg:mod-uzawa-sys} is implemented with $(n_{\max}, k_{\max}) = (20, 200)$ and $\epsilon_S = 10^{-6}$. Once this is algorithm terminates, $\alpha$ is reduced by a factor of $10$ and the final solution, $D^{\text{inv}}$, is taken as an initial guess $D^0$ to rerun Alg. \ref{alg:mod-uzawa-sys}. This process is repeated until the final error $||S-S^{\text{obs}}||_2$, is either no longer decreasing or $||S-S^{\text{obs}}||_2 < \epsilon_S$. Typically this occurs at around $\alpha = 10^{-4}$.
Fig. \ref{fig:d-inv-examples} shows the recovered $D^{\text{inv}}$ for 3 cases of bedrock of with the same slip: $(b_1^2, \beta_2^2)$, $(b_2^2, \beta_2^2)$, and $(b_3^2, \beta_2^2)$. Table \ref{tab:d_err} gives the relative $L2-$norm error between the exact parameter $D^*$ and the recovery $D^{\text{inv}}$ as calculated by eq. \eqref{eq:err_calc} for 12 distinct pairings of bedrock and basal slip.
\begin{table}[h]
	\centering
	\caption{Relative error in $D^{\text{inv}}$, $\mathcal{E}_D$, calculated by eq. \eqref{eq:err_calc}. All values are $\times 10^{-2}$. Colouring indicates size of error (lighter shading is better) and additionally corresponds to the colour of sample solutions in Fig. \ref{fig:d-inv-examples}.}
	\label{tab:d_err}
	\centering
	\includegraphics[width=0.5\textwidth]{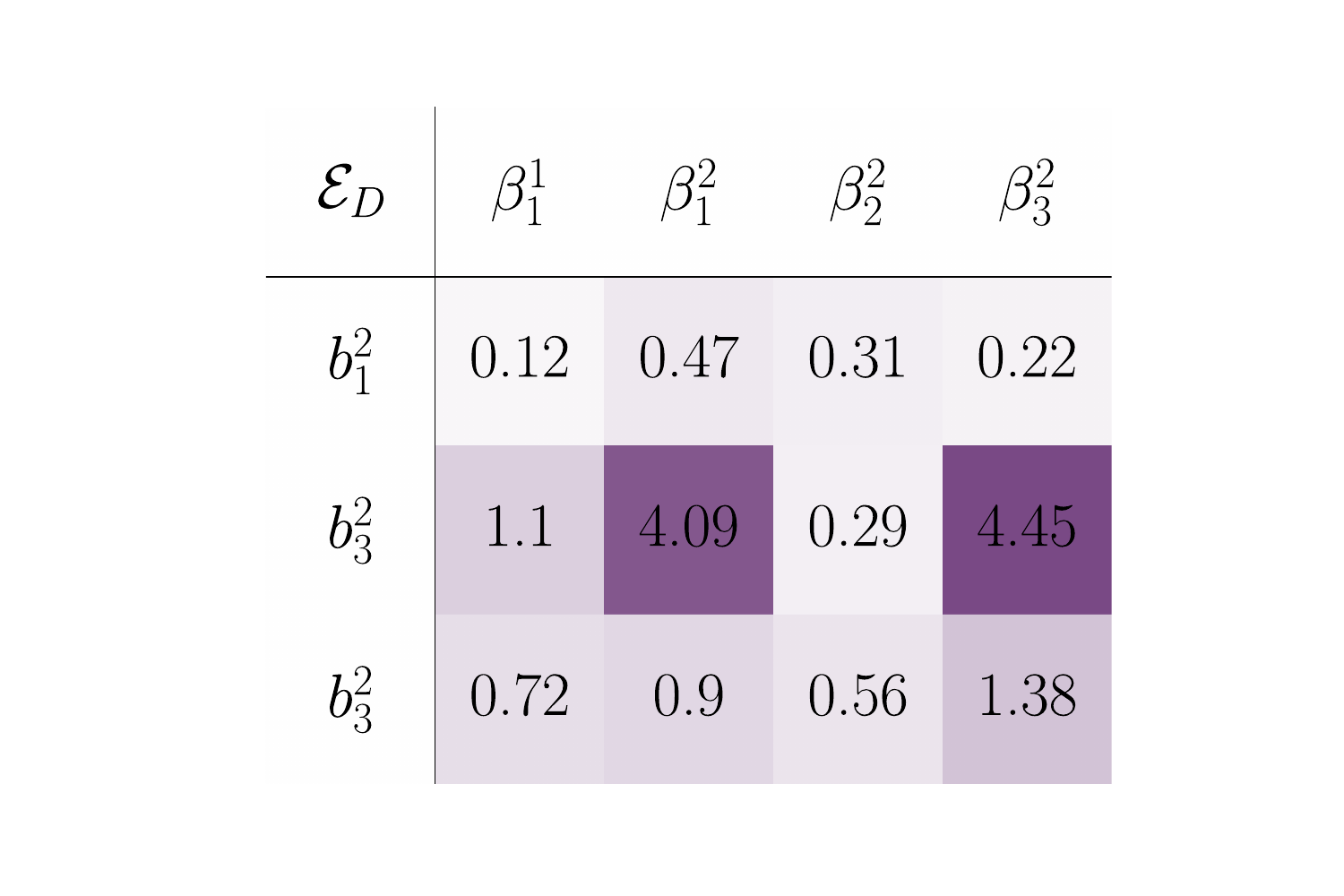}
\end{table}

\subsection{Recovery of $(H, \beta)$}
\label{subsec:h,beta-recover}
Once $D^{\text{inv}}$ is calculated, Alg. \ref{alg:mod-newtons} is implemented to recover $(H^{\text{inv}}, \beta^{\text{inv}})$. Termination criteria for Newton's method is set as $\epsilon_p = 10^{-2}$.

Fig. \ref{fig:h,beta-inv-examples} shows the recovered $H^{\text{inv}}$ alongside the recovered $\beta^{\text{inv}}$ for the same 3 cases of bedrock/slip as shown for $D^{\text{inv}}$: $(b_1^2, \beta_2^2)$, $(b_2^2, \beta_2^2)$, and $(b_3^2, \beta_2^2)$. Table \ref{tab:h,beta_err} (left) gives the relative $L2-$norm error between the exact parameter $H^*$ and the recovery $H^{\text{inv}}$ as calculated by eq. \eqref{eq:err_calc} for 12 distinct pairings of bedrock and basal slip. 

\begin{table}
	\centering
	\caption{Relative error in $(H, \beta)^{\text{inv}}$, $\mathcal{E}_{(H, \beta)}$, calculated by eq. \eqref{eq:err_calc}. All errors are $\times 10^{-2}$. Colouring indicates size of error (lighter shading is better) and additionally corresponds to the colour of sample solutions in Fig. \ref{fig:h,beta-inv-examples}.}
	\label{tab:h,beta_err}
	\centering
	\includegraphics[width=0.49\textwidth]{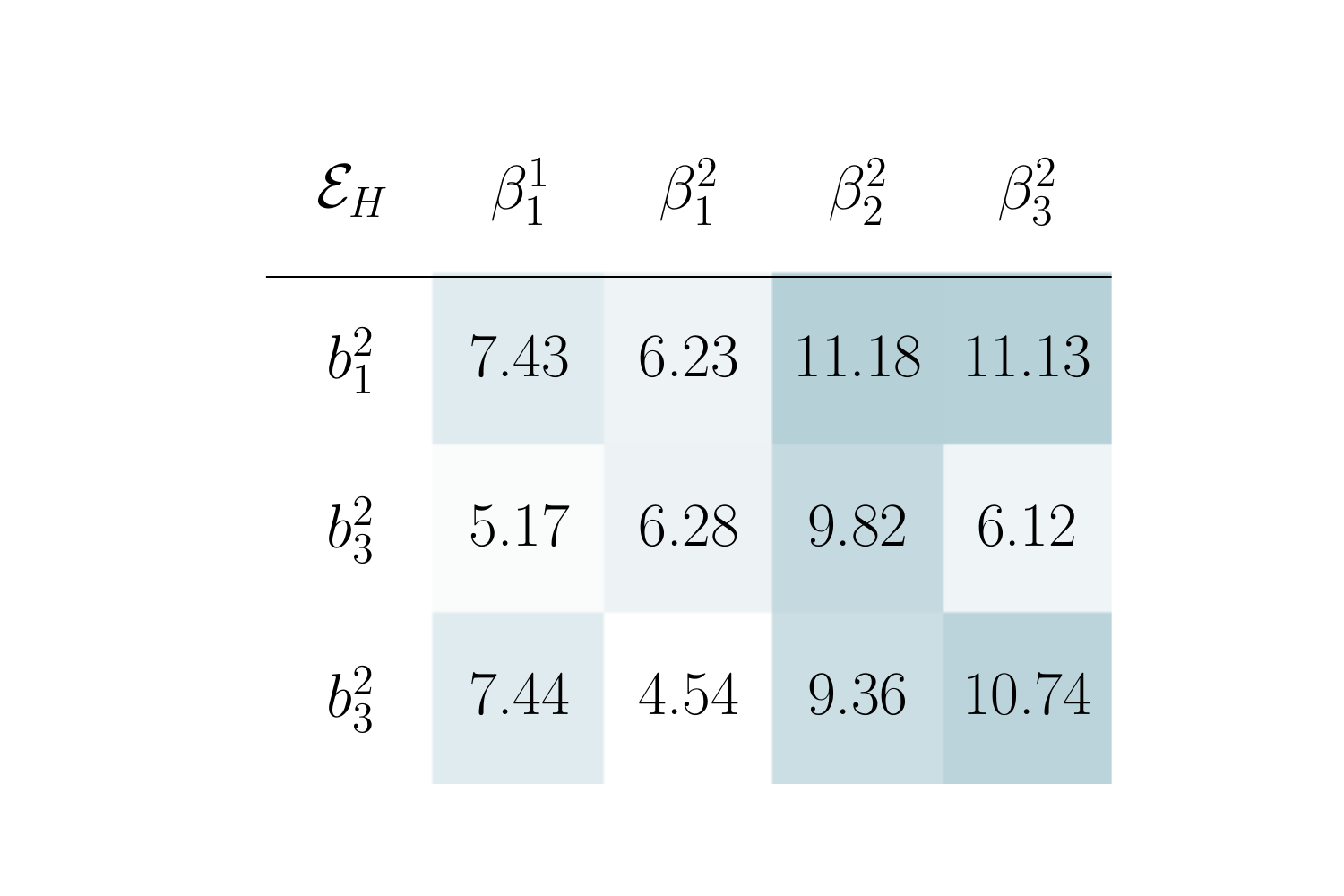}
	\includegraphics[width=0.49\textwidth]{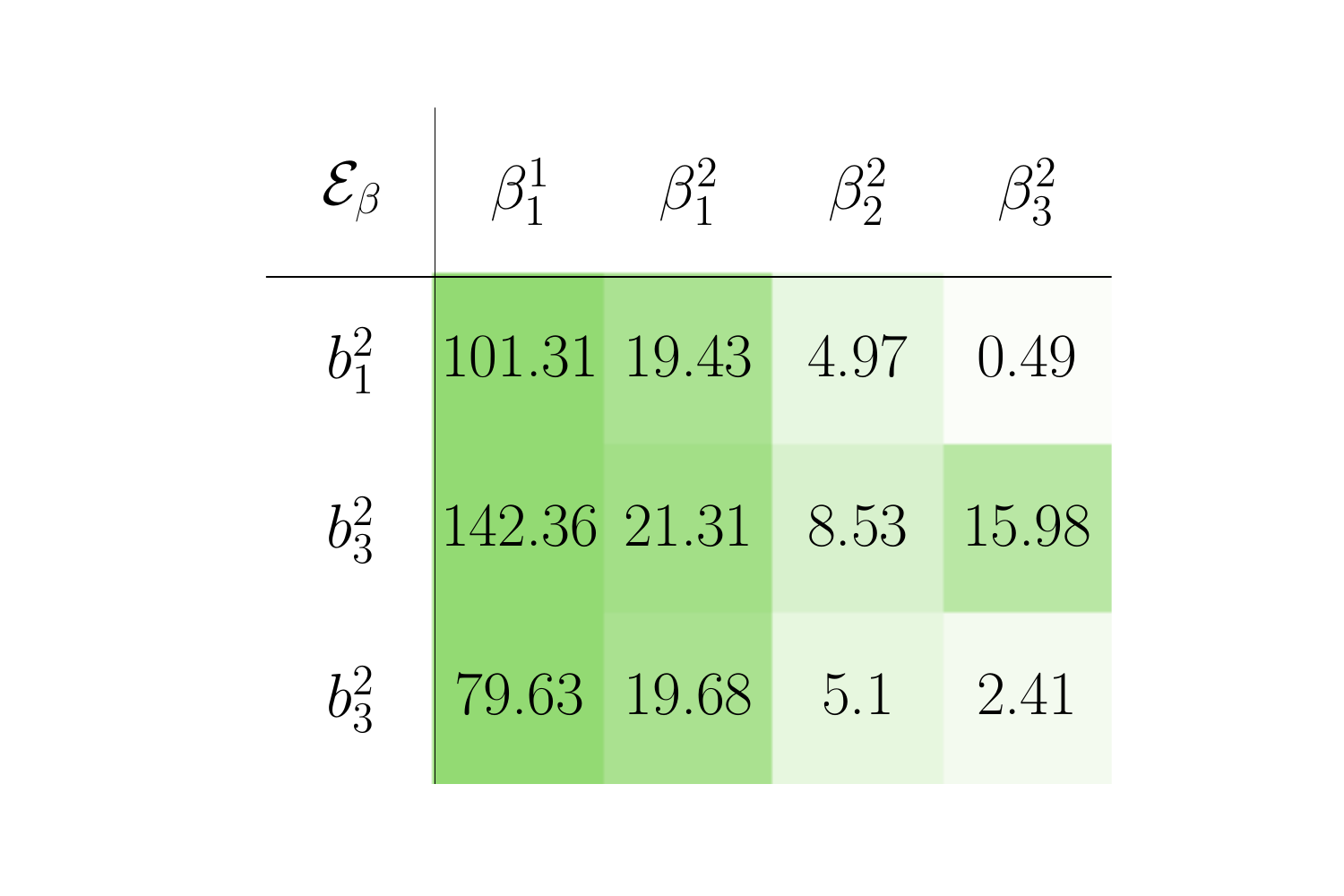}
\end{table}

\begin{figure}[ht]
	\centering
	\includegraphics[width=\textwidth]{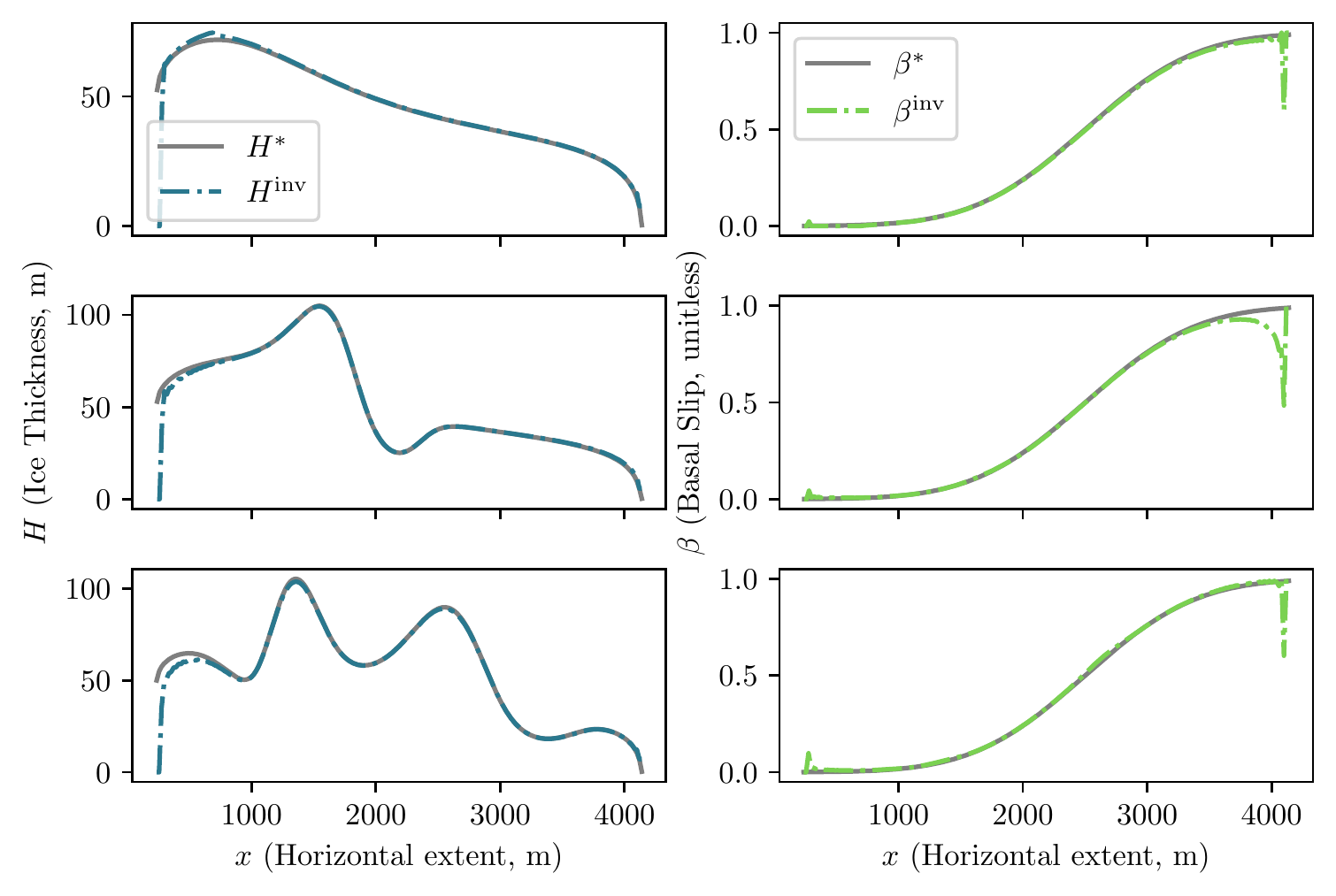}
	\caption{A selection of subsequent pairs $(H^{\mathrm{inv}}, \beta^{\text{inv}})$ using Alg. \ref{alg:mod-newtons}. In all cases, the target bedrock is defined by $b \equiv b_i^2$ ($i \in \{1,2,3\}$) and $\beta \equiv \beta_2^2$.}
	\label{fig:h,beta-inv-examples}
\end{figure}
\subsection{Sensitivity to noise in data.}
\label{subsec:results_noise}
To evaluate the robustness of the proposed methodology, random noise is added to the synthetic data to simulate noise in ice surface measurements and in estimations of the accumulation function. Noise is synthesised in the following way;
\begin{align}
F^\delta &= F^{\text{obs}} + (1+r_\delta)||F^{\text{obs}}||_2, \qquad F \in \{S, f, u_s\} \\
r_\delta & \sim N(0, \delta^2), \qquad \delta \in [0, 1],
\end{align}
where, $\delta$, dictates the amount of noise added to the measurement. Before passing the noisy data into the algorithms, it is filtered as would be done is realistic applications. Here, this is simply done using a moving average with a window width of $200$m. 

Average relative errors in diffusion recovery for 100 random samples noisy data, $S^{\delta}$ and $f^{\delta}$, are given in Tab. \ref{tab:err_noise}. Similarly, relative errors for subsequent ice thickness calculation with 50 samples of noisy surface speed, $u_s^{\delta}$, are also given in Tab. \ref{tab:err_noise} (assuming $D^{\text{inv}}$ calculated with no noise on $S$ and $f$).

\begin{figure}
	\centering
	\includegraphics[width=\textwidth]{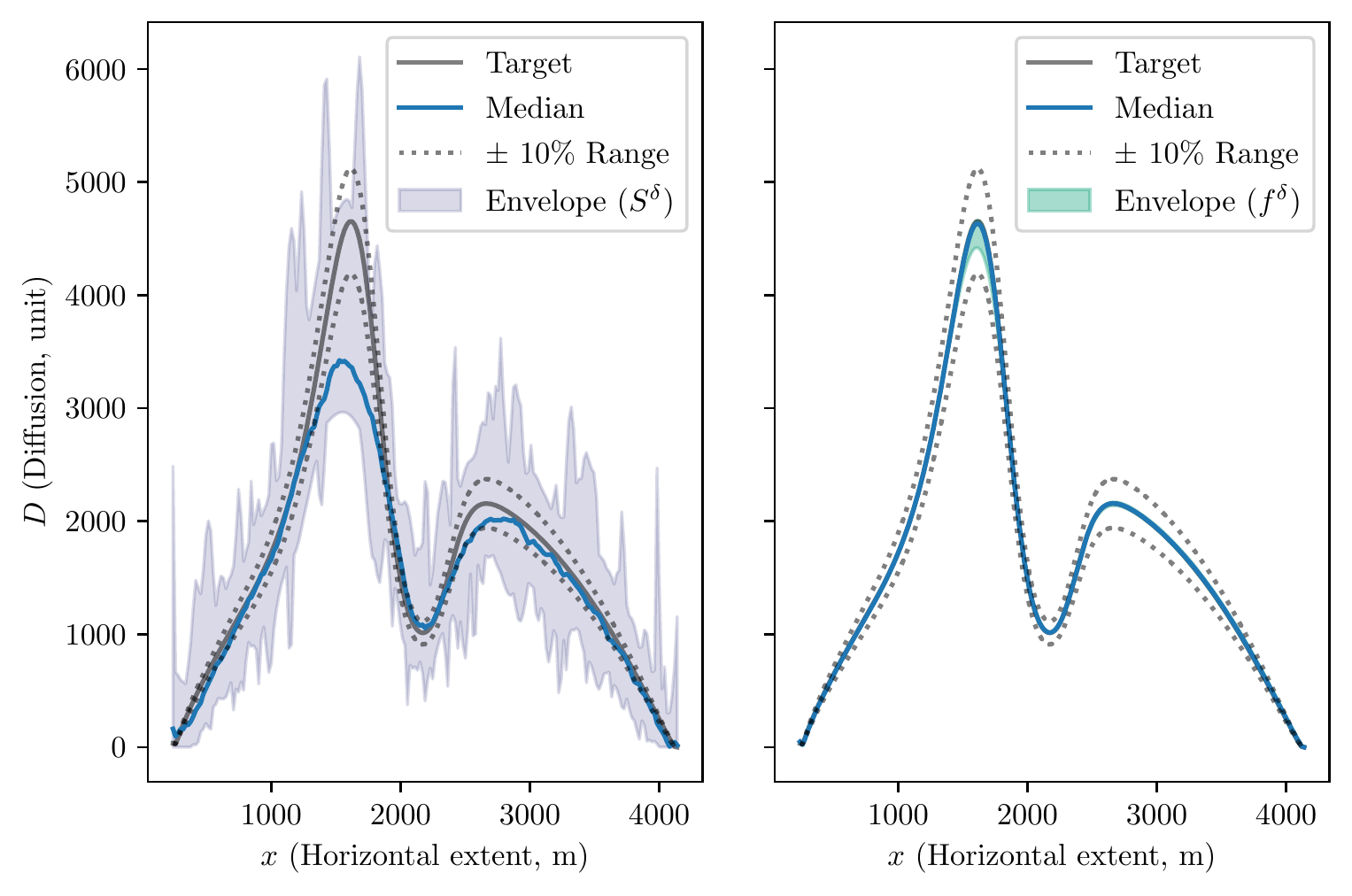}
	\caption{Envelope of 50 samples of recovered diffusion, $D^{\text{inv}}$ for noisy input surface elevation, $S^\delta$, wand accumulation, $f^\delta$. For both, $\delta = 0.05$. Case is $(b_2^2, \beta_2^2)$.}
	\label{fig:envelope-D}
\end{figure}

\begin{figure}
	\centering
	\includegraphics[width=0.7\textwidth]{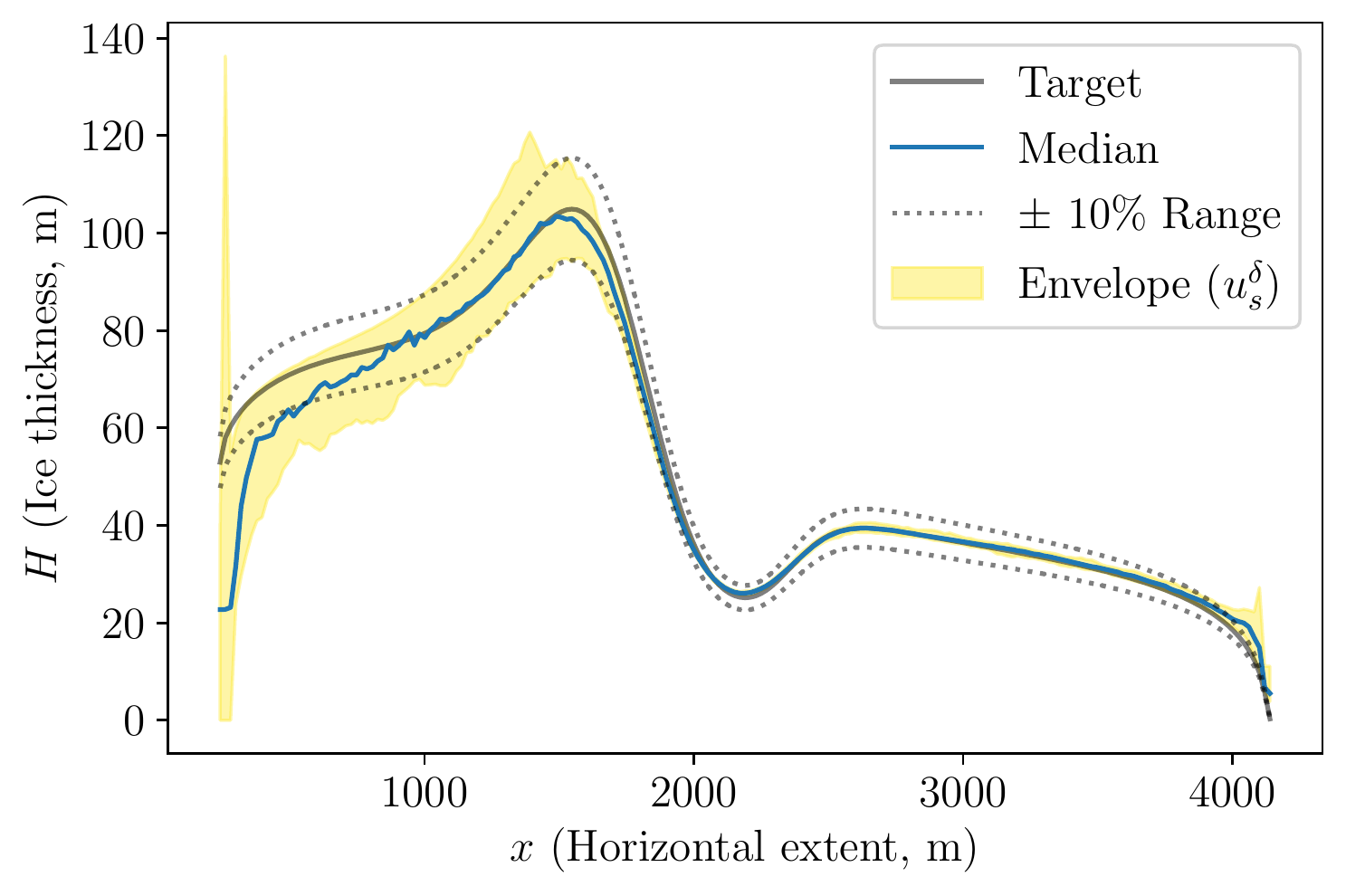}
	\caption{Envelope of 50 samples of recovered ice thicknesses, $H^{\text{inv}}$, for noisy input surface speed, $u_s^\delta$, with $\delta = 0.05$. Case is $(b_2^2, \beta_2^2)$.}
	\label{fig:envelope-H}
\end{figure}

In Fig. \ref{fig:envelope-D}, the target solution, $D^*$, is plotted together with the solution envelopes for both $S^\delta$ and $f^\delta$. Each envelope of solutions is calculated by taking the minimum and maximum solution for $D^{\text{inv}}$ at each coordinate. This plot uses the same 50 samples as in Tab. \ref{tab:err_noise}. Fig. \ref{fig:envelope-H} shows the target ice thickness, $H^*$ with the solution envelope for input $u_s^\delta$. In both figures, 10\% error bands calculated by taking
\begin{align}
F^* \pm 0.1F^* \qquad F \in \{D, H\}
\end{align}
are shown. Additionally, the median solution is overlaid for each solution set.

\renewcommand{\arraystretch}{1.5}
\begin{table}
	\centering
	\caption{Average relative error, $\bar{\mathcal{E}}$, calculated over 50 samples. For $(S^\delta, f^\delta)$;  $\bar{\mathcal{E}}_D$, and for $u_s^\delta$; $\bar{\mathcal{E}}_H$ are given. Error is calculated by eq. \eqref{eq:err_calc}. Noise is added with $\delta = 0.05$ (5\% noise) for all entries. Colouring indicates size of error (lighter shading is better) and additionally corresponds the inversion envelope colour in Figs. \ref{fig:envelope-D} and \ref{fig:envelope-H}.}
	\label{tab:err_noise}
	\includegraphics[width=0.5\textwidth]{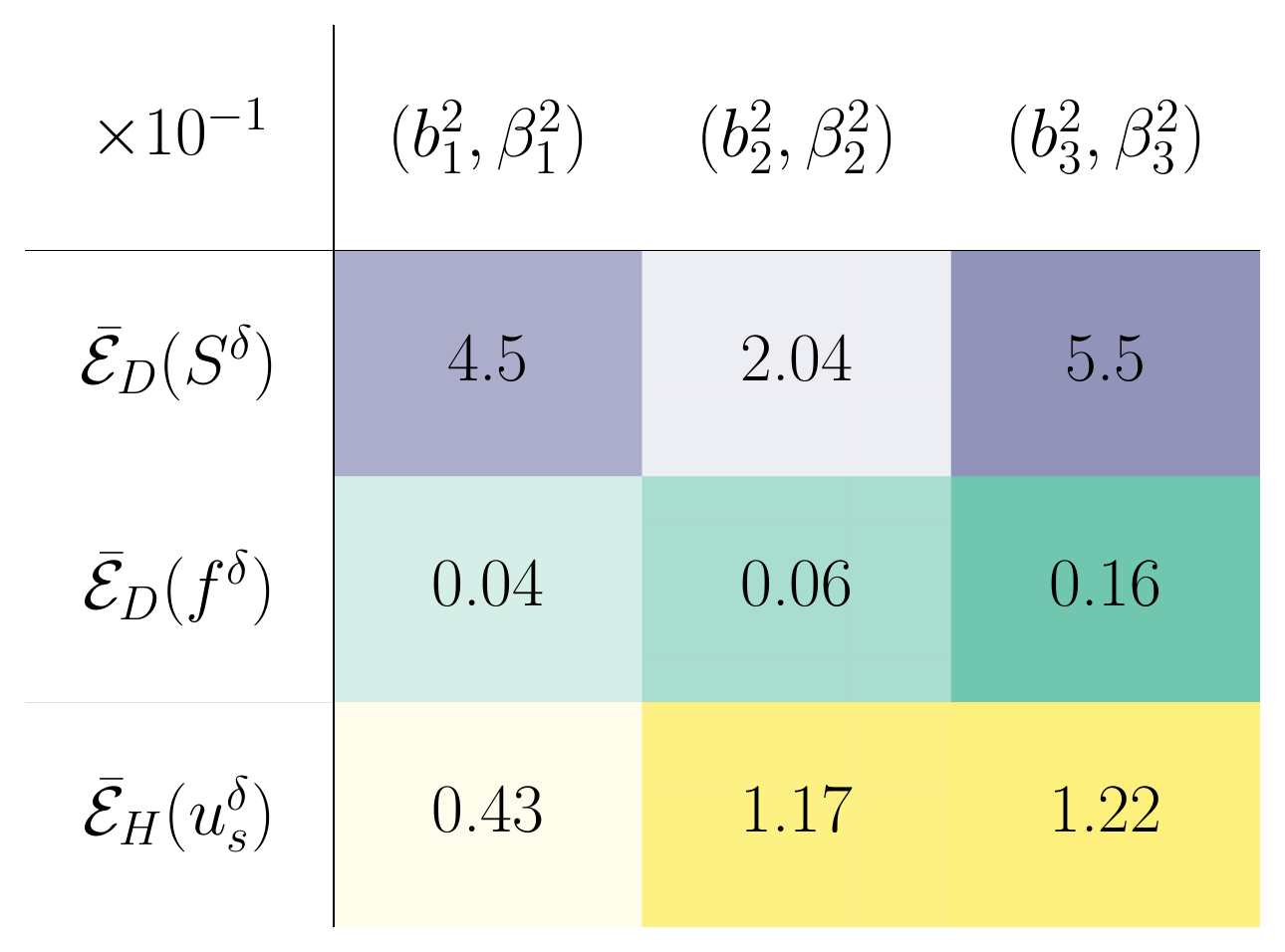}
\end{table}
\renewcommand{\arraystretch}{1}

\section{Discussion}
\label{sec:discussion}

Results presented in Sec. \ref{sec:numerical_results} show that it is possible to use an augmented Lagrangian approach to recover the diffusion in the steady state SIA model from surface elevation data. Accuracy in the recovery without noise was high (with relative error in the magnitude of $10^{-2}$ or less, see Tab. \ref{tab:d_err}) and could subsequently produce good estimations of ice thickness and basal slip (relative error magnitude at most $10^{-1}$ , see Tab. \ref{tab:h,beta_err}) by incorporating additional surface speed data.  

Locations of largest error occur in parts of the domain where SIA assumptions breakdown, for example in locations of steeper underlying bedrock \cite{Bueler2009}. Errors also occur where computations become unstable, for example when $\frac{\partial S}{\partial x} \to 0$ and $u_s \to 0$. This poses difficulty for application of the method to real data as these locations are often of most interest to geo-scientists as they can be the hardest to measure \cite{Morlighem2020}.

Initial sensitivity analysis presented in  Subsec. \ref{subsec:results_noise} indicates that the method can be effective with some noise ($5 \%$) in some surface measurements.

The solution for $D^{\text{inv}}$ is very good (mean errors of magnitude $10^{-2}$) for noise in the accumulation function, $f^\delta$. This is a positive finding as this function can be hard to estimate in practice. For noise of surface elevation, $S^\delta$, the inversion is less successful, with errors of magnitude $5 \times 10^{-1}$. In this case, the method is replicating the bumpiness in the smoothed surface profile by having bumpiness in the solution for $D^{\text{inv}}$. A more advanced method of filtering the surface data could deal with much of this error but further analysis is needed. In the case of noisy surface speed, $u_s^\delta$, results are good for the ice thickness recovery with mean errors of magnitude at most $1.2 \times 10^{-1}$. This is echoed in Fig. \ref{fig:envelope-H}, with the noisy solution envelope falling between the 10\% error bands almost everywhere.

Overall, the sensitivity results are promising for potential future uses of the method with noisy surface measurements as would be the case for field data.

Previous studies either; (a) disregard basal slip \cite{Gessesse2014}, or (b) require prior knowledge of the ice thickness in some locations \cite{McGeorge2021}. The method presented here does not have such limitations.

Overall the method has performed well in the restricted, idealised cases tested in this paper. The main caveats in considering the applicability to real cases are the steady state assumption, the restriction to the unidirectional SIA model, and the wavelength of $\beta$ considered. 

Assumption of a steady state may not be valid for many ice sheets and glaciers. In these cases, a basic work around can be implemented as long as the ice surface is known at two time points, giving $S_1$ and $S_2$, which allows the estimation
\begin{align}
\frac{\partial H}{\partial t} &= \frac{\partial S}{\partial t} \approx \frac{S_2 - S_1}{\Delta t},
\end{align}
and changes the PDE constraint to
\begin{align}
- \nabla \left(D \nabla S \right)= f' = f - \frac{S_2 - S_1}{\Delta t}.
\end{align}
This kind of approximation technique was used successfully by \cite{Gessesse2014}.

Secondly, the unidirectional SIA model is restricted to slow moving grounded ice which restricts the uses for this method. Accounting for limitations in the SIA model itself could be approached from a Bayesian framework as in \citet{Babaniyi2021}. This would be particularly important when considering the confidence of inversions using real data. 

Additionally, the unidirectional nature of the test cases allow for fast computation time. While results here indicate that the same methodology could be applied to a two dimensional case study,  this may require more work as the augmented Lagrangian, while convergent, is computationally inefficient \cite{Fernandez2020}. 

Finally, test cases in this paper all have basal slip distribution with variation over large wavelengths. As \citeauthor{Gudmundsson2008} \cite{Gudmundsson2008} found, small amplitude perturbations in basal slip could only be detected in the surface measurements if the perturbation had a large wavelength in comparison with the ice thickness. If the wavelength was too small, mixing occurred in the surface data between basal slip and bedrock topography which could cause the inverse method to fail in basal slip recovery. This restricts the ability of the inverse method to detect small wavelength changes in basal slip which are physically realistic for many ice flows.

\section{Conclusion}
\label{sec:conclusion}

Overall, the findings presented in this paper reinforce that it is possible to recover both ice thickness and basal slip from surface data.
The method performed well in all test cases showing that it is robust regardless of underlying bedrock or basal slip.
Solutions were good for noisy measurements in the accumulation function and surface speed. The relatively poor solution for noisy surface elevation indicate that this measurement is one of the most important for accuracy which can help to inform scientists in the field.
This is a key result when considering the applicability of the method to `real world' problems in which bedrock and basal slip are unlikely to be uniform or perfectly measured.

Many previous authors have focused on bedrock recovery in no-slip cases, or have recovered bedrock with basal slip by having some prior measurements of ice thickness. This method requires no such prior knowledge making it powerful comparatively.
Additionally, in studies where basal slip is included, methods to date have been complex in comparison. 
The relatively simple method presented here can accurately predict ice thickness and basal slip distribution for certain broadly realistic synthetic cases.

\section*{Acknowledgements}
This work was supported by the University of Canterbury Doctoral Scholarship and the Edward and Isabel Kidson Scholarship.

\printbibliography

\end{document}